\documentclass[12pt, letterpaper]{extarticle}
\usepackage{hyperref} 
\usepackage{xcolor}
\usepackage{soul}
\usepackage{makecell}
\usepackage{url}
\usepackage{adjustbox}
\usepackage[flushleft]{threeparttable}
\usepackage{booktabs}
\usepackage{graphics}
\usepackage{helvet}
\usepackage{times}
\usepackage{comment}
\usepackage{placeins}
\usepackage{multirow}
\usepackage{rotating}
\usepackage{mathtools}
\usepackage{subfig}
\usepackage{arydshln}
\usepackage{longtable}
\usepackage{float}
\usepackage{multicol}
\usepackage{supertabular}
\usepackage{booktabs}
\usepackage{xcolor}
\usepackage{longtable}
\usepackage[framemethod=TikZ]{mdframed}
\usepackage{fullpage}
\usepackage[switch]{lineno}
\usepackage{amsmath}
\usepackage{amssymb}
\usepackage{rotating}
\usepackage{array}
\usepackage{mathtools}
\usepackage[ruled]{algorithm2e}
\usepackage{algorithmic}
\usepackage{bm}
\usepackage{breqn}
\usepackage{comment}
\usepackage{enumitem}
\usepackage{graphicx}
\usepackage{caption}
\usepackage{latexsym}
\usepackage{mathrsfs}
\usepackage{morefloats}
\usepackage{nicefrac}
\usepackage{authblk}
\usepackage{ulem}
\usepackage{booktabs}
\usepackage{array}
\usepackage{pdfpages}
\definecolor{normgreen}{rgb}{0.0, 0.5, 0.0}

\title{Real-time Facial Communication Restores Cooperation After Defection in Social Dilemmas}

\author[1,*]{Mayada Oudah}
\author[1,2]{ John Wooders}

\affil[1]{\normalsize Social Science Division, New York University Abu Dhabi, UAE.}
\affil[2]{\normalsize Center for Behavioral Institutional Design, New York University Abu Dhabi, UAE.}
\affil[*]{\footnotesize Correspondence should be addressed to mayada@nyu.edu}

\date{}

\begin{document} 
\maketitle
\begin{abstract}
\noindent
Facial expressions are central to human interaction, yet their role in strategic decision-making has received limited attention. We investigate how real-time facial communication influences cooperation in repeated social dilemmas. In a laboratory experiment, participants play a repeated Prisoner's Dilemma game under two conditions: in one, they observe their counterpart's facial expressions via gender-neutral avatars, and in the other no facial cues are available. Using state-of-the-art biometric technology to capture and display emotions in real-time, we find that facial communication significantly increases overall cooperation and, notably, promotes cooperation following defection. This restorative effect suggests that facial expressions help participants interpret defections less harshly, fostering forgiveness and the resumption of cooperation. While past actions remain the strongest predictor of behavior, our findings highlight the communicative power of facial expressions in shaping strategic outcomes. These results offer practical insights for designing emotionally responsive virtual agents and digital platforms that sustain cooperation in the absence of physical presence. 

\vspace{0.25in}

\end{abstract}

\section*{Introduction}

Facial expressions play a critical role in shaping human interactions. As a fundamental channel of nonverbal communication, facial expressions convey emotions and intentions without the need for words, yet significantly influence others' behavior and expectations~\cite{Ekman, role2009, knapp, Sagliano2022, Kleef2022}. They guide social behavior, evoke empathy, and help regulate cooperation and conflict in both casual and strategic settings such as negotiations, team decision making, and everyday interactions~\cite{Hatfield1993, Holland_2020, Kleef_2009, CENTORRINO20158}. The ability to recognize and interpret facial expressions in such settings allows individuals to better anticipate reactions and adjust their decisions accordingly~\cite{Ekman2003}. 

Despite its central role, much of the economic literature on strategic decision-making has abstracted away from nonverbal communication~\cite{sally2000general, camerer2003behavioral}. In particular, the canonical Prisoner's Dilemma and its repeated versions typically do not account for facial signals, even though such cues are omnipresent in real-world social dilemmas~\cite{axelrod1984evolution, fehr2000fairness, binmore2007playing}. Understanding the role of facial expressions in strategic interactions is therefore not only scientifically relevant but also practically important in a digital world where in-person encounters are increasingly replaced by text-based communication.

Recent advancements in biometric technology now allow us to investigate the importance of facial expressions with precision. In this study, we utilize state-of-the-art technology to automatically and accurately measure and classify facial expressions in real time, via a webcam, as individuals interact. An individual's captured expression is represented to their counterpart by an avatar. 

 Crucially, this methodology overcomes a major hurdle in behavioral analysis: the signal-to-noise problem inherent in human faces. In standard interactions, the ``signal" of an emotion is often confounded by the ``noise" of the sender's identity—where attributes like race, gender, and attractiveness can overpower the expression itself and form biases~\cite{facialcues2010}. By re-targeting biometric data to gender-neutral avatars, we effectively isolate the emotional signal from these identity cues . We design and evaluate a novel set of gender-neutral avatars to accurately represent the universal set of facial expressions as defined by Ekman~\cite{Ekman}.

In prior research on the role of emotions in social dilemmas \cite{Reuben2009, Mussel2013, deMelo2015, deMelo20}, by contrast, participants select a facial expression or a text description of their emotional state from a predefined set of faces or states, either before or after making a decision. In this case, participants may behave strategically in conveying their emotional state. We address this challenge in our study by capturing and communicating facial expressions in real time as participants interact with each other.

In our experiment, pairs of participants play the repeated Prisoner's Dilemma game for at least 30 rounds, with a continuation probability of 90\% thereafter. We study how the ability to observe facial expressions -- communicated in real-time via avatars -- affects cooperation rates. In the Facial Communication (FC) treatment, players see their counterpart's facial expressions as represented by a gender-neutral avatar that reflects the most intense emotion detected by the biometric software. In the control treatment, no Facial Communication (nFC), players receive no visual emotional feedback. 

Our results show that, in addition to sustaining high mutual cooperation rates, the ability to observe facial expressions significantly increases cooperation following defection. This suggests that visual social cues soften the perceived severity of non-cooperative behavior, encouraging participants to respond with greater forgiveness. These findings highlight the power of embedding real-time emotional feedback in social dilemmas, where facial expressions play a restorative role in maintaining cooperative dynamics. 

\section*{Results}
\begin{figure}[H]
\centering
\includegraphics[width=0.9\textwidth]{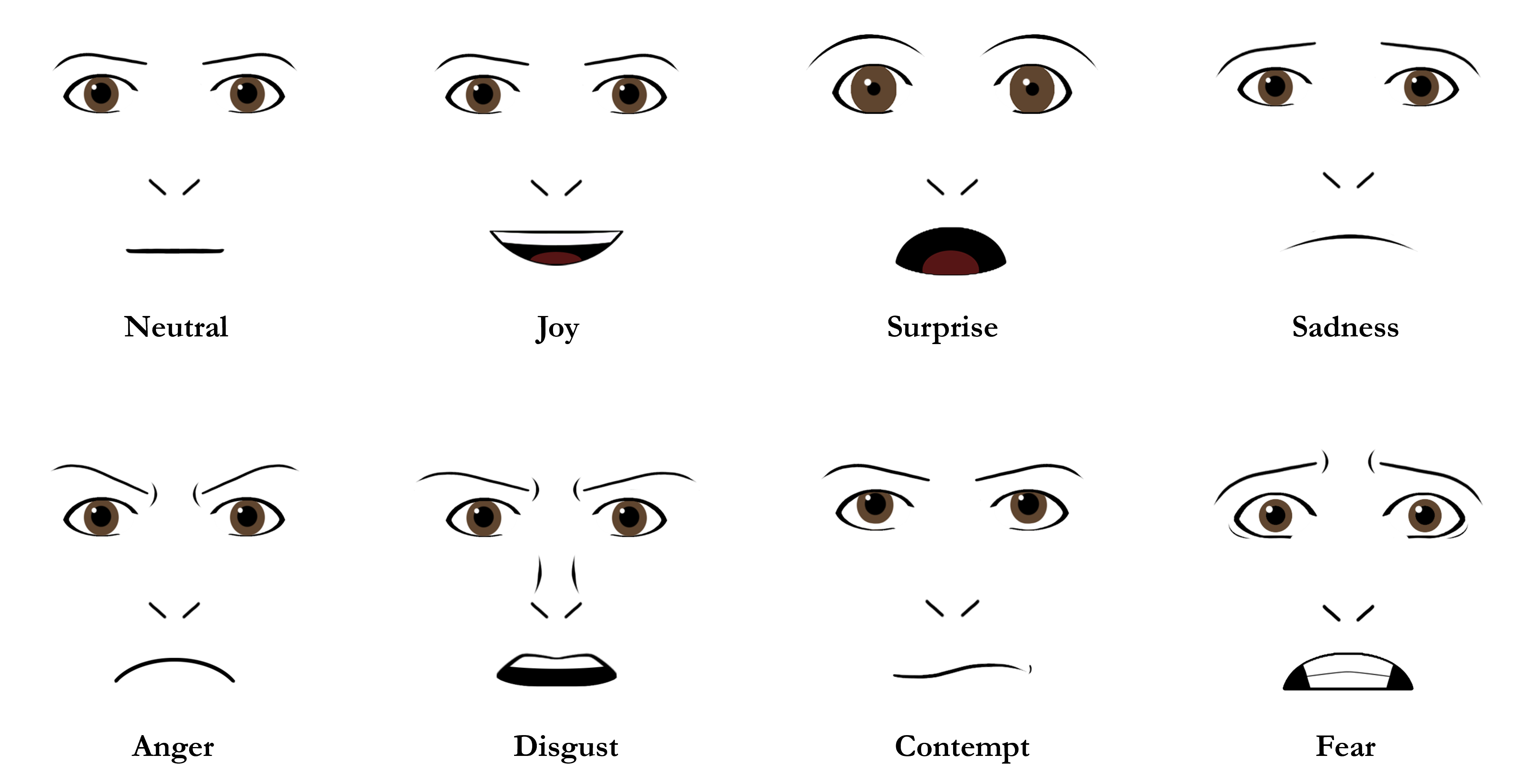}
\caption{\textbf{The avatars used to represent facial expressions.} The avatars utilize a schematic design to minimize the ``Uncanny Valley" effect while maximizing signal clarity.
}
\label{fig:avatars}
\end{figure}

\begin{figure}[H]
\centering
\includegraphics[width=0.9\textwidth]{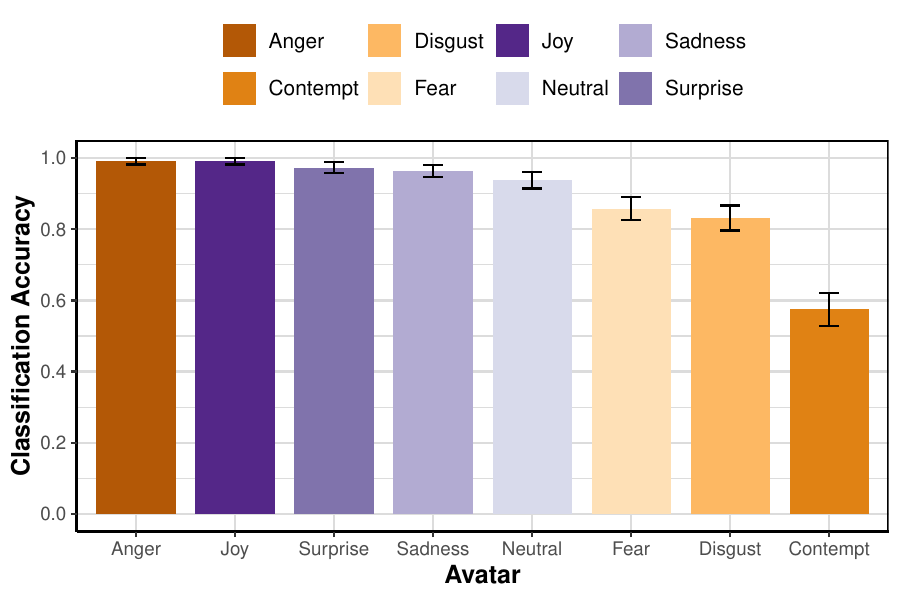}
\caption{\textbf{Avatars classification accuracy.} Error bars show the standard error of the mean.}
\label{fig:evaluation}
\end{figure}

\begin{figure}[H]
\centering
\includegraphics[width=0.9\textwidth]{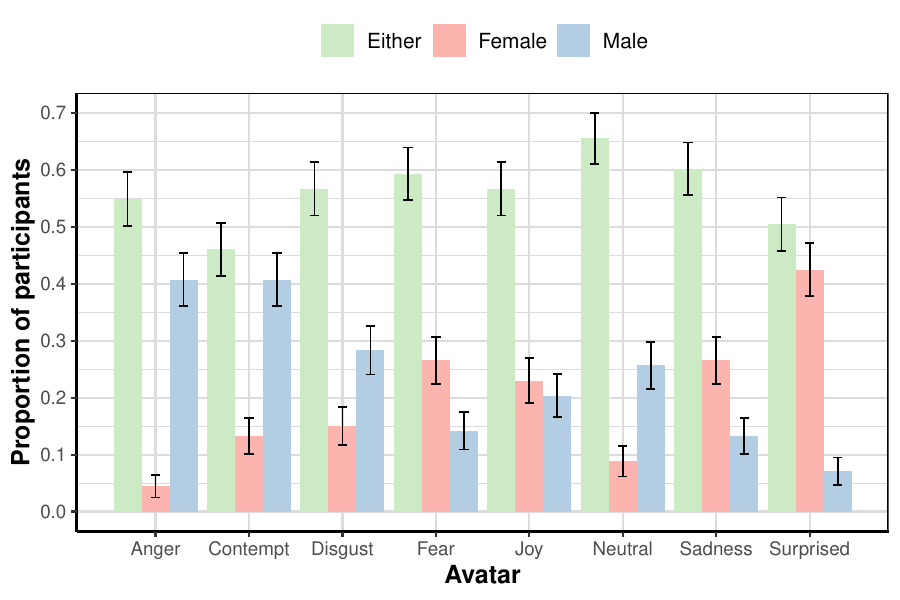}
\caption{\textbf{Avatars gender classification.} Error bars show the standard error of the mean.}
\label{fig:gender}
\end{figure}

To evaluate the avatars we designed, shown in Figure~\ref{fig:avatars},  we invite subjects to participate in a Qualtrics survey. In the survey we ask each participant to classify the avatars into 9 categories, consisting of the 8 universal emotions (Joy, Sadness, Surprise, Fear, Anger, Disgust, Contempt, Neutral) and the category ``Hard to Classify." We also ask participants to classify the avatars by gender (Male, Female, Can be Either) by dragging an image of the avatar and dropping it into the category where they believe it belongs. Figure~\ref{fig:evaluation} shows the participants' classification accuracy. All of the avatars achieved a classification accuracy rate greater than 80\%, with the exception of \textit{Contempt} (accuracy $\approx$ 57\%). This lower accuracy is consistent with established literature, which documents that contempt is frequently conflated with Disgust or Anger and is difficult to recognize~\cite{contempt}. Indeed, the majority of misclassifications for this avatar fell into ``Anger", ``Disgust", and ``Hard to Classify" categories (see Supplementary information Figure~S1), confirming that the avatar successfully captured the intended negative valence despite the semantic ambiguity. Figure~\ref{fig:gender} shows the participants' gender classification. For each avatar except Contempt, the majority of participants assigned ``Can be Either" as the gender classification.

Given the high classification accuracy of the avatars (except for Contempt) and that ``Can be Either" is the most frequent gender classification, we use these avatars to represent participants' facial expressions in the Prisoner's Dilemma experiment. The experiment has two treatments that differ in whether a player observes the avatar of his counterpart. We label the treatment in which facial communication is enabled as FC and the treatment in which facial communication is disabled as nFC (for further details, see the Methods section). In the FC treatment, facial expressions are communicated via an avatar. Each player observes, in real-time on their own screen, the facial expression of the other player.

\begin{figure}[H]
\centering
\includegraphics[width=0.9\textwidth]{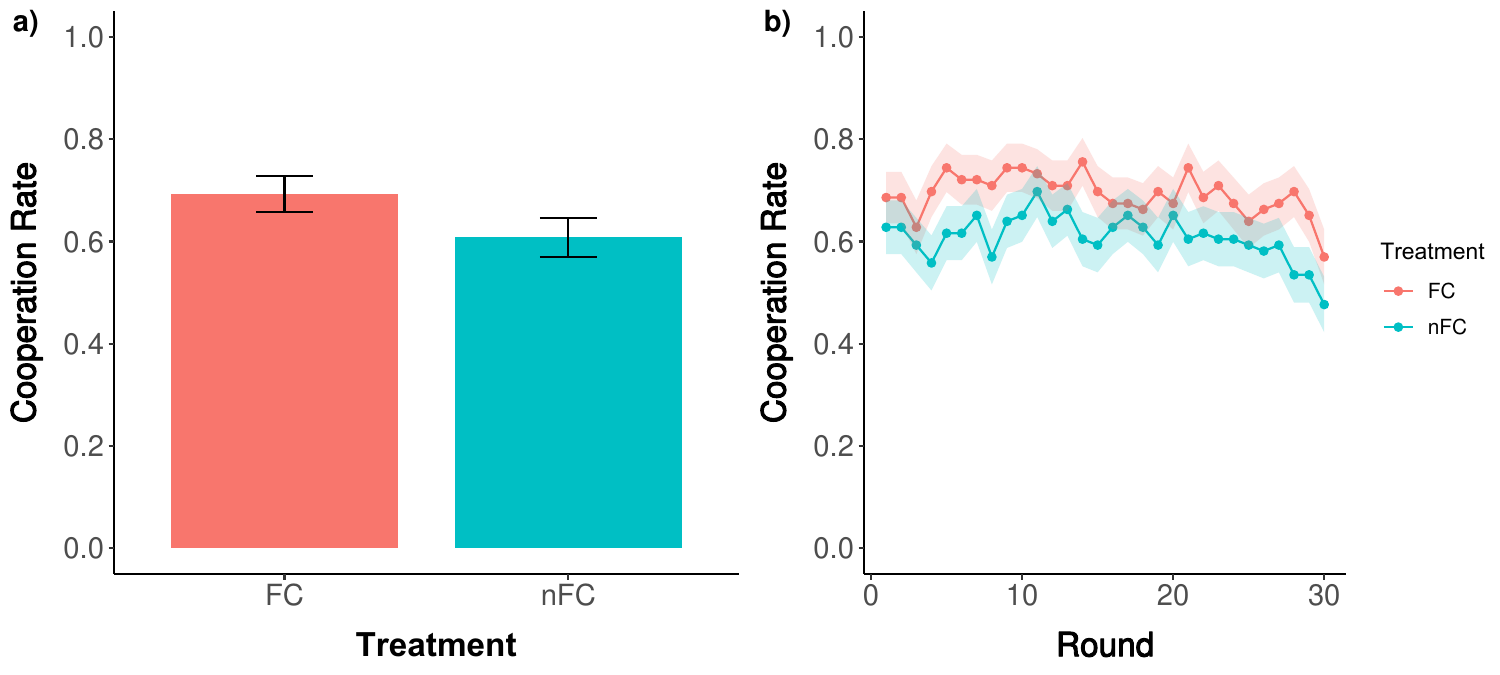}
\caption{\textbf{Experimental results -- Cooperation.} \textbf{a)} Shows the cooperation rate in each treatment. Error bars show the standard error of the mean. \textbf{b)} Depicts cooperation rates across 30 rounds.
}
\label{fig:coop_results}
\end{figure}

Figure~\ref{fig:coop_results} shows the cooperation rates for both treatments. Participants cooperate more in the FC treatment than in the nFC treatment. This result is validated through regression analysis, using the Generalized Linear Mixed Model (GLMM) to account for within-subject dependencies and repeated measurement. Figure~\ref{fig:coef_1} visualizes the regression coefficients (see Supplementary Table S1 for full model details). The positive coefficient for FC ($\beta = 0.47, p < 0.001$) indicates that the presence of facial avatars significantly increases cooperation rates compared to the anonymous baseline.

To understand the influence of perceived facial expressions on the decision to cooperate, we regress whether or not a participant cooperated on (i) own and other's displayed facial expression prior to making a decision and (ii) the cooperation decisions of both players in the prior round. Figure~\ref{fig:coef_2} shows that while the presence of Joy (Own and Other) is a positive predictor of cooperation, recent history of actions (Own and Partner's Previous Cooperation) is the strongest predictor of current behavior, confirming the dominance of reciprocal strategies such as Tit-for-Tat. See Supplementary Table S2 for full model details

\vspace*{-4mm}
We run further analysis to investigate whether participants' facial expressions are influenced by their own and the counterpart's actions. Crucially, this analysis relies on the raw biometric values captured by the iMotions software, which differs from the facial expressions displayed to the counterpart. Unlike the avatar, which applied a 60\% intensity threshold to ensure signal clarity, these measured values capture subtle emotional shifts and internal reactions that may not have been visible to the other player.

Figure~\ref{fig:emo_own} shows how participants' own actions influence these internal emotional states. It is evident that participants are more emotionally engaged (expressing more non-Neutral emotions) in the FC treatment than in the nFC treatment. When participants cooperate, they tend to express positive emotions -- especially Joy -- suggesting emotional satisfaction or prosocial signaling (see Figure~\ref{fig:emo_own} (a) and (c)). This aligns with the view that cooperation is emotionally rewarding.

In the FC treatment, own defection is associated with less Joy and a rise in negative emotions (see Figure~\ref{fig:emo_own} (b)). This biometric spike potentially indicates internal conflict, guilt, or social disapproval—affective reactions that occur even if the expression was too subtle to trigger the avatar. In contrast, in the nFC treatment, own defection is not associated with any significant change in the defector's measured emotional state, highlighting the lower emotional stakes of anonymous interaction.

Figure~\ref{fig:emo_other} illustrates the influence of the counterpart's actions on the participant's measured internal emotional state. Consistent with the findings on own actions, participants in the FC treatment exhibit significantly higher emotional reactivity than those in the nFC treatment. 

When the counterpart cooperates, participants respond with a distinct rise in Joy, reflecting social approval, gratitude, or a shared prosocial orientation (see Figure~\ref{fig:emo_other} (a) and (c)). Conversely, when the counterpart defects, it triggers strong negative biometric signals in the FC treatment—specifically spikes in Contempt and Sadness (see Figure~\ref{fig:emo_other} (b)). This sharp negative reaction (absent in the anonymous condition) suggests that when facial identity is present, defection is perceived not merely as a strategic loss, but as a social or moral violation.

Crucially, the timeline of emotional states reveals a difference in expectations. In the nFC treatment, ``pre-Joy" (measured Joy before the outcome is revealed) is significantly lower in rounds where the counterpart eventually defects (see Figure~\ref{fig:emo_other} (d)). This indicates that in anonymous play, defection is often the status quo; participants are trapped in a cycle of mutual defection and negative affect, correctly anticipating further betrayal.

\begin{figure}[H]
\centering
\includegraphics[width=0.8\textwidth]{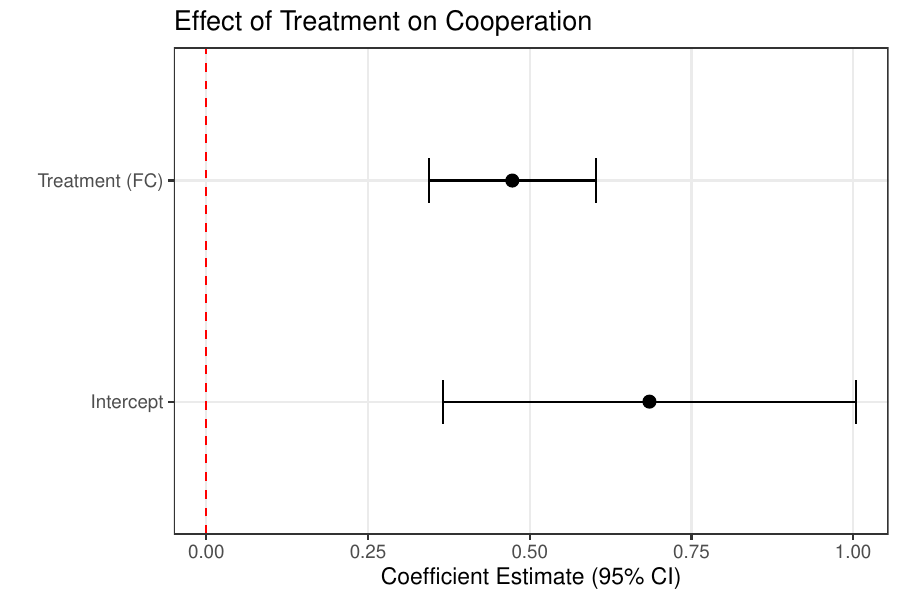}
\caption{\textbf{Facial communication significantly increases the likelihood of cooperation.}
Fixed-effect coefficient from a GLMM predicting cooperation ($N=5,160$). The point estimate represents log-odds relative to the nFC baseline; error bars denote 95\% confidence intervals.}
\label{fig:coef_1}
\end{figure}

\begin{figure}[H]
\centering
\includegraphics[width=0.8\textwidth]{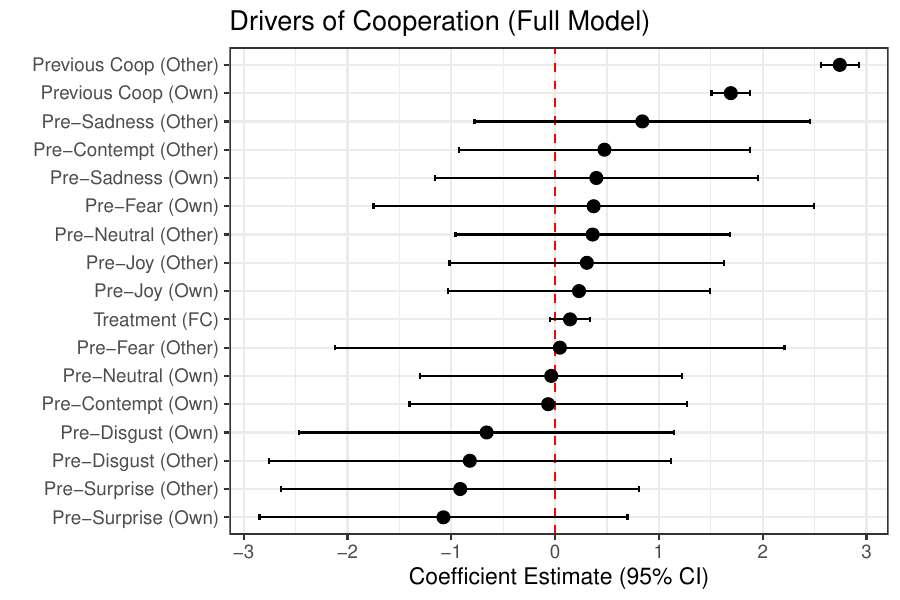}
\caption{\textbf{History of play is the strongest predictor of cooperation.} GLMM estimates comparing structural and emotional variables. Points represent log-odds; error bars denote 95\% confidence intervals.}
\label{fig:coef_2}
\end{figure}

\begin{figure}[H]
\centering
\includegraphics[width=0.8\textwidth]{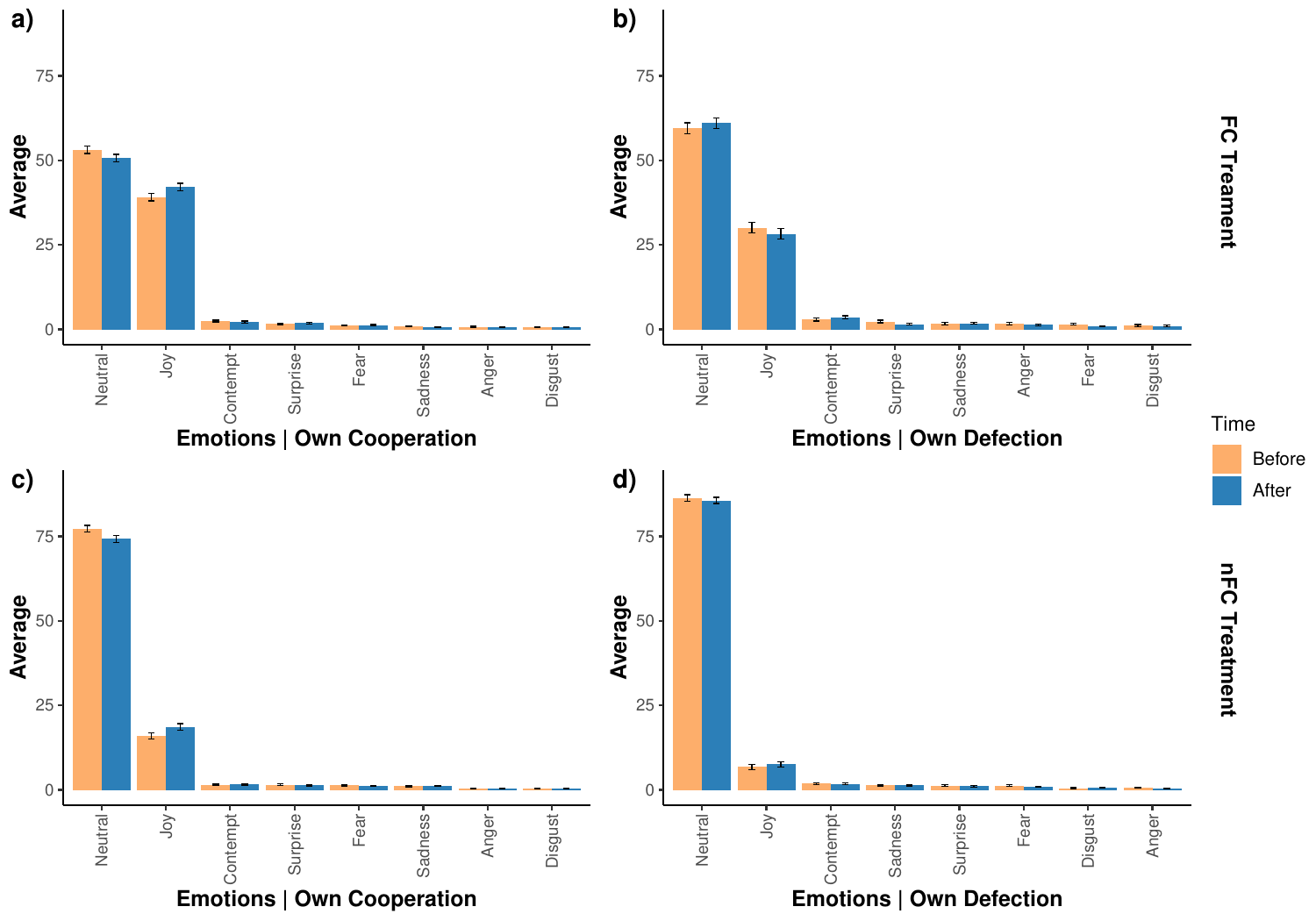}
\caption{\textbf{Measured Emotions Before and After Own Action.} Error bars show the standard error of the mean.}
\label{fig:emo_own}

\vspace*{4mm}

\includegraphics[width=0.8\textwidth]{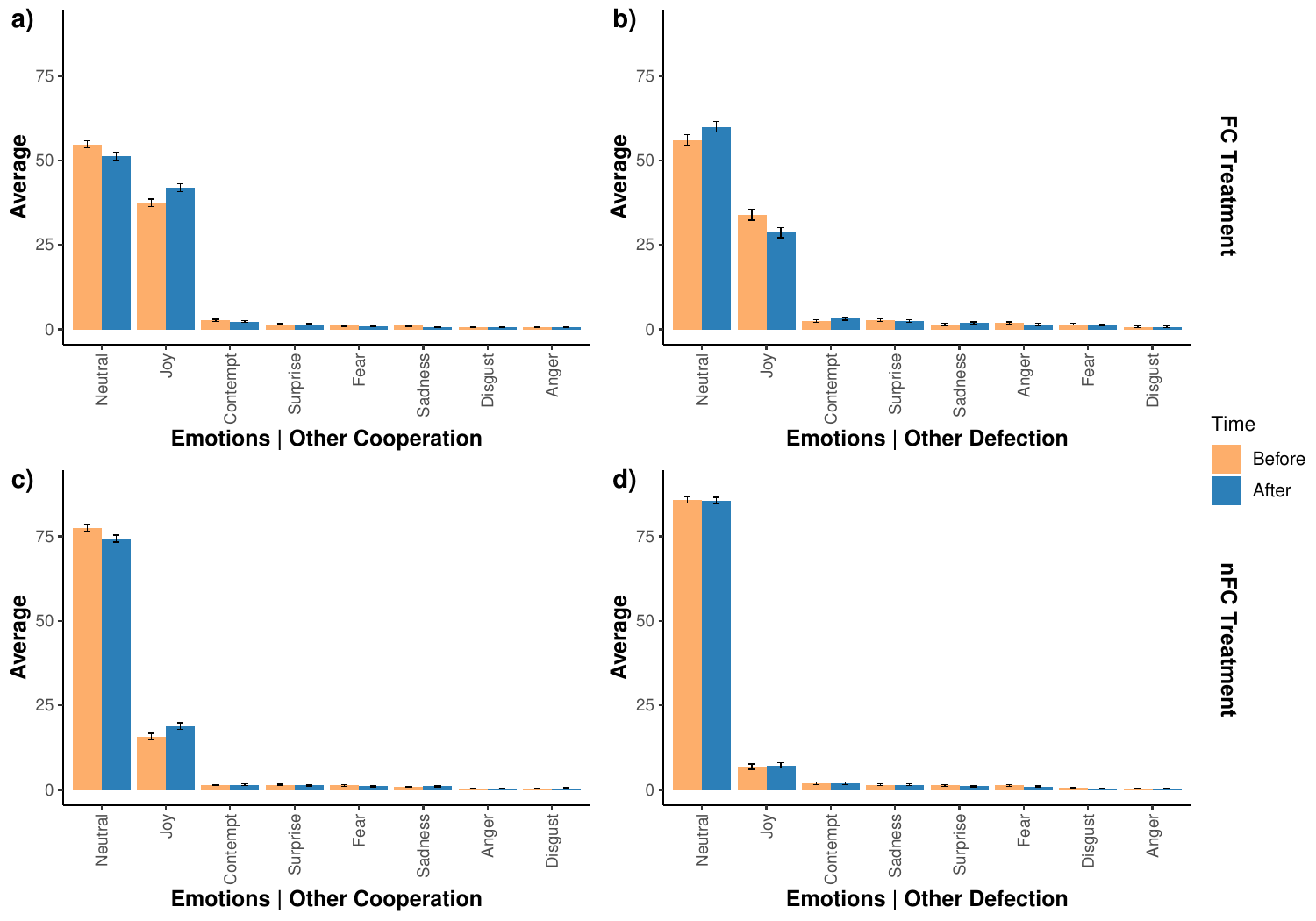}
\caption{\textbf{Measured Emotions Before and After The Other Player's Action.} Error bars show the standard error of the mean.}
\label{fig:emo_other}
\end{figure}

In contrast, in the FC treatment, pre-Joy remains high even in rounds where the counterpart subsequently defects. This suggests that FC participants maintain a state of emotional optimism or cooperative intent right up until the moment of betrayal. The defection in the FC treatment therefore acts as a violation of this sustained hope, explaining the intense restorative efforts (forgiveness) that follow. Further information can be found in our Supplementary information.

\begin{figure}[H]
\centering
\includegraphics[width=\textwidth]{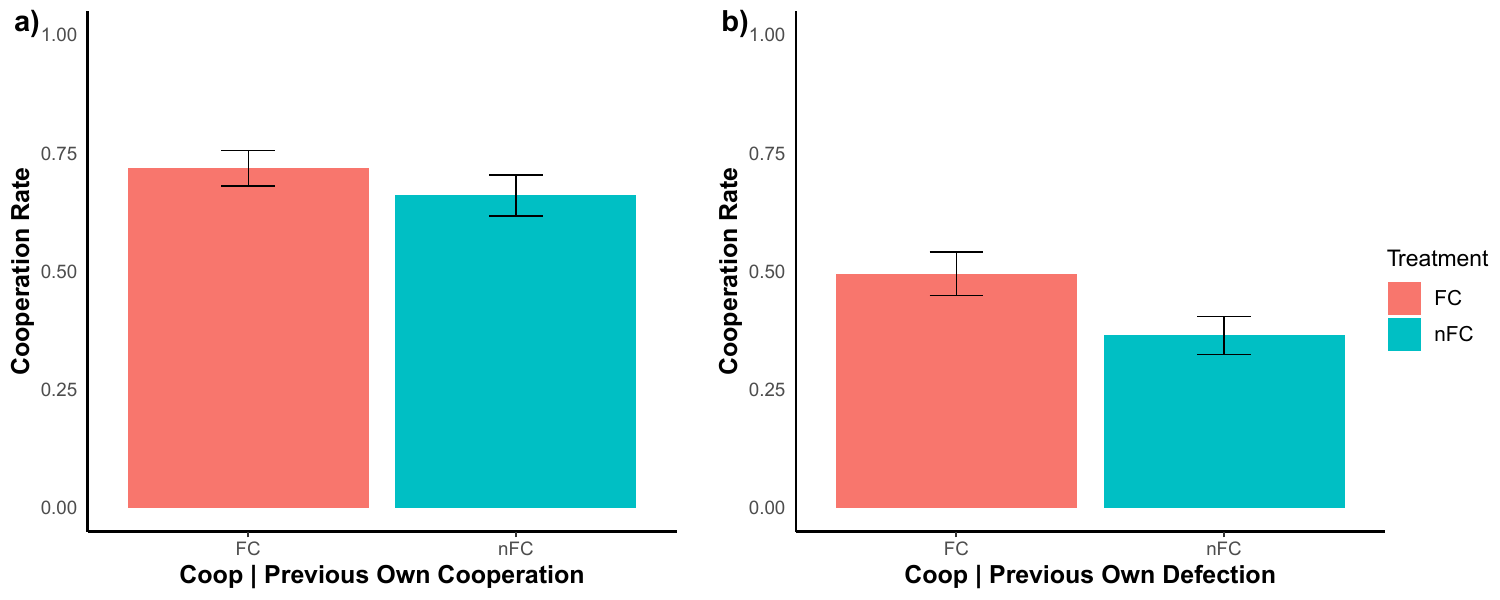}
\caption{\textbf{Conditional Cooperation of Own Previous Decision.} \textbf{a)} Cooperation rate given own cooperation in the previous round and \textbf{b)} Cooperation rate given own defection in the previous round. Error bars show the standard error of the mean.}
\label{fig:cc_own}
\end{figure}

\begin{figure}[H]
\centering
\includegraphics[width=\textwidth]{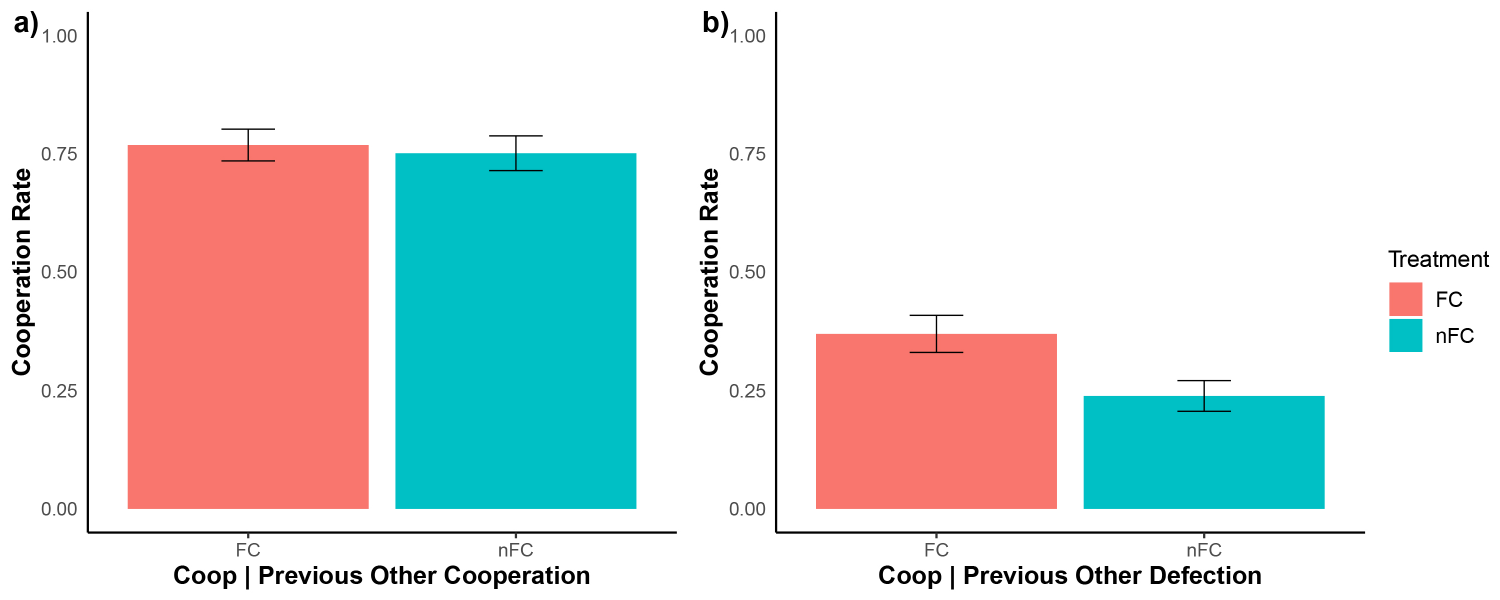}
\caption{\textbf{Conditional Cooperation of the Other Player's Previous Decision.} \textbf{a)} Cooperation rate given the other player's cooperation in the previous round and  \textbf{b)} Cooperation rate given the other player's defection in the previous round. Error bars show the standard error of the mean.}
\label{fig:cc_other}
\end{figure}

Figure~\ref{fig:cc_own} (a) shows the participants' cooperation rate given that they cooperated in the previous round, for both FC and nFC treatments. Likewise, Figure~\ref{fig:cc_other} (a) shows the participants'  cooperation rate given that their counterpart cooperated in the previous round, for both treatments. In each case, there is no statistically significant difference in cooperation rates between the treatments. 

There is, however, a striking difference between the treatments if a player defected in the previous round.  Figure~\ref{fig:cc_own} (b) shows that participants are significantly more likely to cooperate in the FC than in the nFC treatment following their own defection in the previous round. Participants may feel a sense of guilt and regret following their own defection when facial expressions are communicated between players, that they don't feel in the absence of facial communication. Figure~\ref{fig:cc_other} (b) shows that participants are also significantly more likely to cooperate in the FC than in the nFC treatment following their counterparts' defection. In the presence of facial communication, participants are more forgiving and more cooperative after their counterparts' defection. This suggests that visual social cues (facial expressions) play a restorative role in maintaining cooperative dynamics.

At the end of the experiment, each participant completed a survey to rate the other player's likability, intelligence, cooperativeness, trustworthiness, selfishness, forgivingess, tendency to bully, and whether they would like to interact with them again (see Figure~\ref{fig:forgiveness} (a) and the Supplementary material for more details). We find a statistically significant difference in the reported ``Forgivingess" of the other player between the two treatments. FC participants perceive the other player to be more forgiving than nFC participants (see Figure~\ref{fig:forgiveness} (b)). This result supports that Facial Communication promotes forgiveness following a defection -- see Figure~\ref{fig:cc_other} (b).

\section*{Discussion}
Our results show that face-to-face communication via avatars promotes cooperation in social dilemmas, fundamentally changing the perception of a defection. In the nFC treatment, a defection is just a cold numerical choice and the standard tit-for-tat logic (``you defect, I defect") dominates. But when the participant can see their counterpart's avatar before and after their counterpart defects, they obtain extra social and emotional information that lets them handle their betrayal with more forgiveness. That is, in the FC treatment, one might pick up on an expression -- e.g., Sadness following defection and Joy following cooperation. Those facial cues signal that the defection may have been unintentional or that the defector feels bad about it, which makes the participant more willing to give them another chance.  By contrast, in the nFC treatment, the participants see the outcome without any avatar. There is no socially costly apology‐signal to soften the blow, and thus they retaliate more than in the FC treatment. 

The effect of facial communication is especially pronounced after a defection, whether one's own or the counterpart's. In standard repeated games, defection typically initiates a downward spiral, with reciprocity mechanisms punishing defections. Here, however, facial expressions appear to soften the rigidity of that response. Players in the FC treatment are more likely to re-establish cooperation after it has been breached, indicating that facial cues provide an interpretive context to actions that would otherwise be perceived as purely self-interested and hostile.

Facial expressions function as affective ``repair tools," helping participants to attribute remorse or alternative motives to a defection that might otherwise be viewed as intentional. This is consistent with work showing that emotions -- particularly those expressed facially -- are not just reflective but communicative; they convey intent, invite empathy, and facilitate reconciliation \cite{Sagliano2022, Kleef2022, deMelo2015}.

While the history of play establishes the baseline for cooperation, our analysis reveals that facial expressions function as a critical 'circuit breaker' specifically when that baseline is fractured. Although past actions remain the primary driver of routine behavior -- consistent with standard reciprocity -- emotional cues become decisive following a violation. By recontextualizing an unexpected defection, facial expressions modulate the rigid logic of tit-for-tat, allowing players to prioritize forgiveness over retaliation. Thus, while memory sustains stability, emotional communication restores it.

Importantly, players in the FC treatment not only behaved more cooperatively but also were more emotionally engaged. Their facial expressions, especially Joy and Contempt, were more pronounced and responsive to both their own and others' actions. This increased expressiveness suggests that being seen, and seeing others, increases the social salience of the interaction and promotes cooperation.

These findings carry several practical implications. In digital settings, where facial feedback is usually absent, basic avatar-driven emotional expressions can revive elements of accountability, empathy, and shared understanding. For designers and developers of virtual agents and AI systems, our results highlight the importance of embedding real-time emotional feedback loops that are interpretable and responsive, not merely reactive.

\begin{figure}[H]
\centering
\includegraphics[width=1\textwidth]{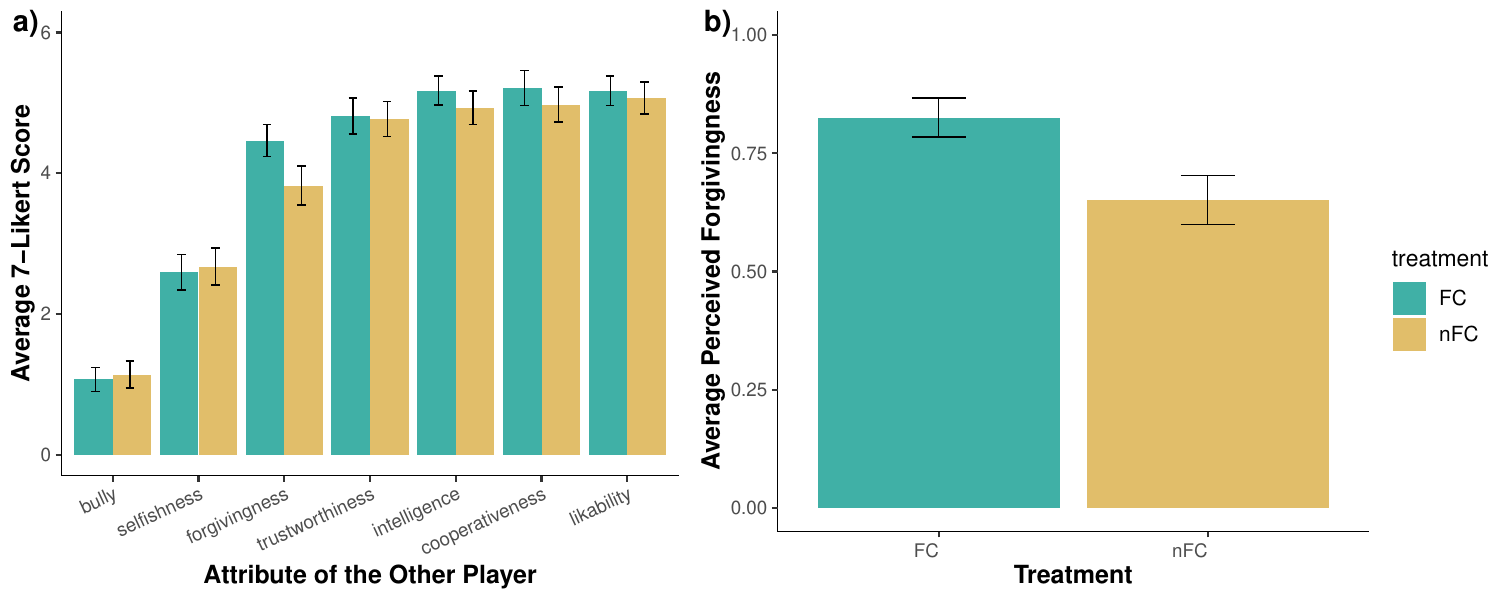}
\caption{\textbf{Participants' perception of the other player's attributes.} \textbf{a)} The average perceived scores for each attribute of the other player. \textbf{b)} The frequency that participants report a Likert score of 3 or above of the other player's forgivingess. Error bars show the standard error of the mean.}
\label{fig:forgiveness}
\end{figure}

Our experimental design involves specific trade-offs inherent to the digital mediation of human affect. While our real-time emotion mapping ensures strict experimental control, the imposition of a 60\% intensity threshold likely filters out subtle micro-expressions. By suppressing low-intensity signals (e.g., a faint smile below the threshold), our design prioritizes signal clarity over sensitivity to prevent avatar flickering or signal instability. Consequently, our results may represent a conservative estimate of the restorative power of facial expressions, as subtle conciliatory cues were effectively censored. Despite the artificial nature of the avatars, the high ratings for Likability in Figure~\ref{fig:forgiveness} (a) show that participants treated the interaction as a genuine social exchange, not just a human-computer interaction task.

All put together, our findings suggest that even subtle facial communication, in controlled conditions, can meaningfully shift behavioral outcomes in strategic environments. By transforming how individuals interpret defection, facial expressions provide a window into intent and a platform for emotional repair. Future research could explore how these mechanisms operate across different cultural contexts, group sizes, or in high-stakes environments, as well as investigate whether artificial agents designed to express human-like emotions can similarly foster trust and cooperation. Understanding how emotional visibility shapes long-run dynamics is especially timely in an increasingly digital world -- where the capacity to ``see and be seen" may determine not only what we choose, but who we become in relation to others.

\section*{Methods}
All of our protocols received IRB approval from the NYU Abu Dhabi (NYUAD) Internal Review Board. In the first experiment, the subjects assessed the accuracy of the designed avatars. The second experiment, on the interactive Prisoner's Dilemma, followed a between-subject design, and participants were recruited for different treatments on different days. Participants were not aware of the different treatments. Subjects signed a consent form before participating. 

\subsection*{Experimental design and setup}
To investigate the impact of facial expressions on decision making, we first designed our own set of gender-neutral avatars to represent the universal set of 8 emotions identified by Ekman~\cite{Ekman}: neutral, sadness, happiness, fear, anger, surprise, disgust, and contempt.\footnote{We designed our own avatars since, in the existing literature, avatars are often gendered.} Figure~\ref{fig:avatars} shows the avatars created. The schematic nature of the avatars was a deliberate choice to avoid the ``Uncanny Valley" effect, where hyper-realistic but imperfect avatars cause revulsion~\cite{uncanny, oudah24}. Showing avatars instead of the players' actual faces eliminates potential decision biases towards  facial characteristics such as gender, ethnicity, attractiveness, and so on \cite{Eckel2006, Stirrat2010, CHEN201265}. To verify the accuracy of the designed avatars, in the first experiment subjects evaluate the avatars by classifying them according to the universal emotions and according to gender (Male, Female, Can be Either). 

\begin{figure}[H]
\centering
\includegraphics[width=\textwidth]{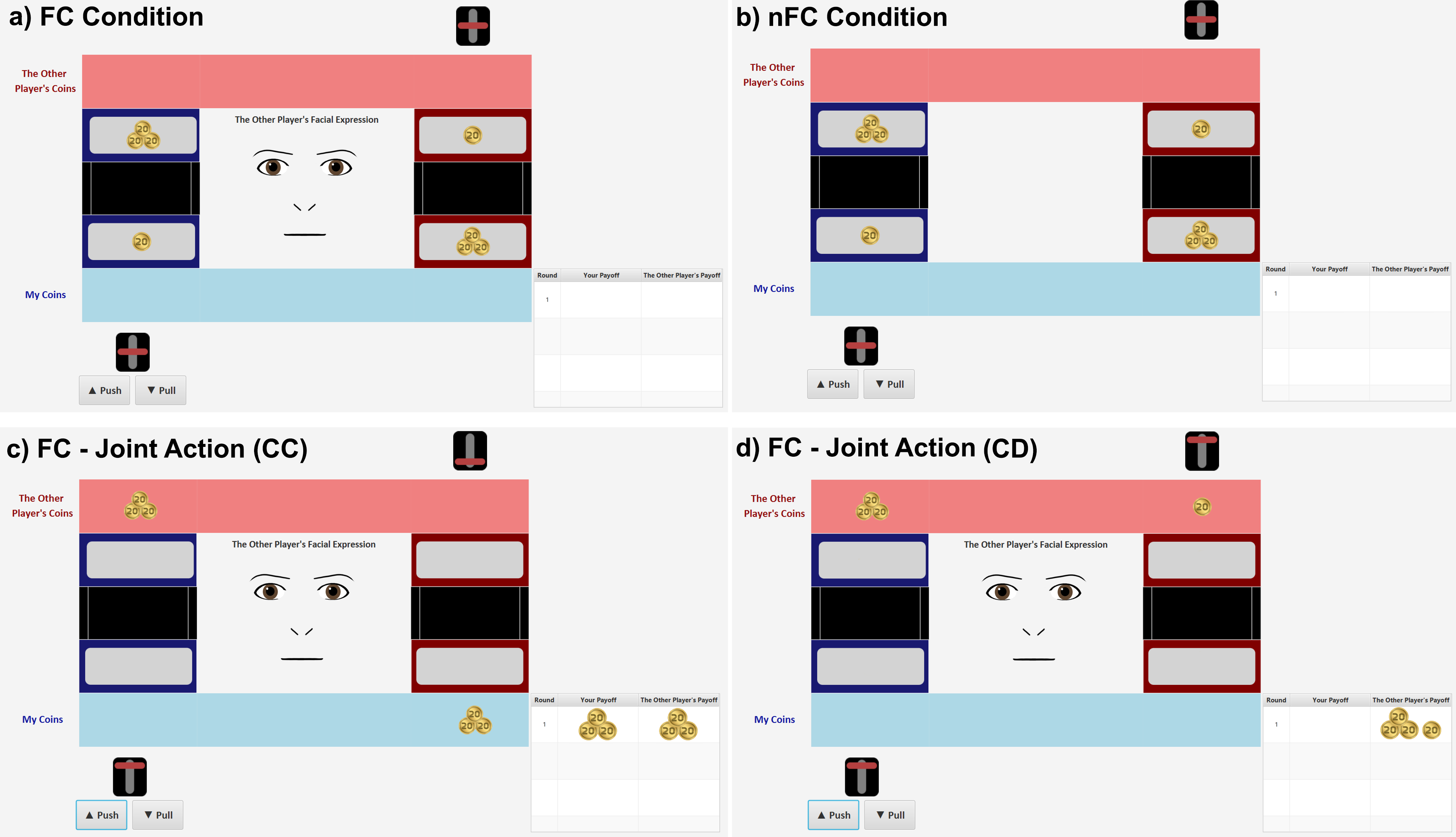}
\caption{\textbf{User interface and treatments.} \textbf{a)} The interface of our 2-player Prisoner's Dilemma game under the FC treatment, where the player can observe the avatar of the other player. Participants control the player on the bottom and see the other player on the opposite side. Each player can either push (Cooperate) or pull (Defect) the lever by clicking the corresponding buttons for these actions. Pushing the lever delivers three coins to the other player and causes one coin to fall into a black hole. Pulling the lever delivers one coin to the player and causes three coins to fall into the black hole. \textbf{b)} The interface under the nFC treatment. The rules are the same as the FC treatment, but the player does not observe the avatar of the other player. \textbf{c)} The interface of mutual cooperation. \textbf{d)} The interface of cooperating against a defector.}
\label{fig:gui}
\end{figure}

We chose to study the Prisoner's Dilemma since it is an engaging game where players might exhibit a broad range of facial expressions. In the experiment, we used the graphical user interface (GUI) of Blake et al.~\cite{Blake2015} (see Figure~\ref{fig:gui}). In the game, two players face each other, with the participant being the player at the bottom and his counterpart being the player on top. Each player can either ``push" or ``pull" their tray by clicking the corresponding button. The game is played simultaneously since a player does not observe their counterpart's action before they take their own. The game is played by fixed pairs.

We build on the original GUI by adding a block in the middle of the screen where the counterpart's facial expression is represented via an avatar. In particular, we display the avatar corresponding to the most intense emotion, as measured by the biometric software, so long as its intensity exceeds a threshold of 60\%; otherwise we display the Neutral avatar. If the software does not detect a face, the block is empty. 

There are two experimental treatments: (i) Facial Communication (FC) and (ii) no Facial Communication (nFC). In both treatments, the participants' facial expressions are captured and measured in real-time using iMotions~\cite{iMotions}, which provides accurate and automatic facial expression analysis using Affectiva Affdex~\cite{Stckli2018FacialEA, iMotionsAccuracy}. The software detects and measures facial expressions via a webcam on top of each monitor. In the FC treatment, participants observe their counterpart's avatar throughout the entire experiment. In the nFC treatment, participants' facial expressions were recorded but not displayed. See the Supplementary material for more details about iMotions platform. 

The experimental sessions were held in the Social Science Experimental Lab (SSEL) at NYUAD, where participants sat at private cubicles. We invited participants in groups of 8 per session. After participants read and signed consent forms, the experimenter reviewed the instructions and participants undertook a tutorial on their computer to familiarize themselves with the mechanics of taking an action in the game and calibrate the software to their facial expressions. Participants were informed that their facial expressions are recorded and analyzed throughout the play of the PD game, and were instructed to avoid covering their face. Participants were informed that they would play the PD game with another fixed participant for at least 30 rounds, after which there would be a 10\% chance that the interaction would terminate at the end of each following round. In this setting, cooperation at each round is sustainable as a subgame perfect equilibrium using strategies like Grim Trigger or Tit-for-Tat. Participants were also informed they would be paid for two random rounds at the end of the experiment.

Once the instructions were presented, the experimenter attended to each participant and ensured they were seated properly for the webcam (Logitech C920 Widescreen HD Pro) to accurately capture their face and they were requested to make several facial expressions to ensure that the iMotions module could detect their facial expressions. After the iMotions calibration was complete, the participants were asked to input their player ID (which was provided to them by the experimenter upon entering the lab), their age, and gender in the game software. Once the information was entered by paired players, the game began. iMotions software was running in the background while the game was active. There was no timer set for participants to make a decision in each round. The game lasted approximately 10--15 minutes, depending on the number of rounds realized and how fast participants made decisions. When the interaction ended, a screen with the total earnings was displayed. Participants filled out a post-experiment survey that had questions related to how they made their decisions in the game, demographics, and how do they rate the other player's behavior. Participants were paid using YouGotaGift, a digital gift card company in the Middle East that supports electronic gift cards of over 700 brands, including Amazon.

\subsection*{Recruitment}
We recruited students from NYUAD, who were 18 years or older, using the ORSEE recruitment system to evaluate the designed avatars and to participate in our experimental studies. To be eligible to participate in the PD game, students could not have a thick beard or wear glasses, in order to avoid distorting the quality of the data collected by iMotions. Students who evaluated the designed avatars were also allowed to participate in the PD game if they satisfied both of these eligibility criteria. A student was not allowed to participate in both the FC and the nFC treatment.

Data collection for the survey took place between the 14$^{\mathrm{th}}$ of October and the 3$^{\mathrm{rd}}$ of November, 2022. For the FC Treatment, it took place between the 29$^{\mathrm{th}}$ of May and the 16$^{\mathrm{th}}$ of June, 2023, and for the nFC treatment, it took place between the 6$^{\mathrm{th}}$ and the 24$^{\mathrm{th}}$ of November, 2023. We recruited 113 participants for the avatar evaluation survey, and a total of 172 participants for the PD game, 86 for the FC treatment and 86 for the nFC treatment.

Participants received AED 15 (\$4.08) for completing the evaluation survey, which took an average completion time of 8.60 minutes. The participants in the PD game received an AED 30 (\$8.17) show-up fee in addition to a bonus that ranged between AED 0 (\$0) and AED 190 (\$51.73) depending on their decisions and the randomly selected rounds for payment. The average compensation in the FC treatment was AED 121.86 (\$33.18) and in the nFC treatment was AED 116.02 (\$31.59). The average completion time was 20 minutes. 

\vspace*{-5mm}

\subsection*{Subject characteristics}
Participants who completed the avatar evaluation survey were 56.64\% female and 42.48\% male, 36.28\% of which come from South Asia (16.81\% from Europe and Central Asia, 14.16\% from East Asia and Pacific, 11.50\% from Middle East and North Africa, 10.62\% from Sub-Saharan Africa, 6.19\% from Latin America and Caribbean, and 3.54\% from North America), with an average age of 20.97 (sd = 1.38). Participants who completed the FC experimental treatment were 60.47\% male and 39.53\% female, with an average age of 20.31 (sd = 1.51). Participants who completed the nFC experimental treatment were 52.33\% male, 46.51\% female, and 1.16\% other. The participants' average age was 20.27 (sd = 1.46). 

\vspace*{-5mm}

\section*{Data availability} 
All data analyzed in these studies will be deposited on Open Science Framework at the time of publication and will be made available for review upon request.

\section*{Code availability} 
The plots, regressions, and statistical analysis were executed using R (version: 4.3.2). The code will be deposited on Open Science Framework at the time of publication and will be made available for review upon request.
\newpage
\bibliography{naturebib.bib}

@book{camerer2003behavioral,
title     = {Behavioral Game Theory: Experiments in Strategic Interaction},
author    = {Camerer, Colin F.},
year      = {2003},
publisher = {Princeton University Press},
address   = {Princeton, NJ}
}

@article{sally2000general,
title   = {A General Theory of Sympathy, Mind-Reading, and Social Interaction, with an Application to the Prisoners' Dilemma},
author  = {Sally, David},
journal = {Social Science Information},
volume  = {39},
number  = {4},
pages   = {567--634},
year    = {2000},
doi     = {10.1177/0539018000394004}
}

@book{axelrod1984evolution,
title     = {The Evolution of Cooperation},
author    = {Axelrod, Robert},
year      = {1984},
publisher = {Basic Books},
address   = {New York}
}

@article{fehr2000fairness,
title   = {Fairness and Retaliation: The Economics of Reciprocity},
author  = {Fehr, Ernst and Gächter, Simon},
journal = {Journal of Economic Perspectives},
volume  = {14},
number  = {3},
pages   = {159--181},
year    = {2000},
doi     = {10.1257/jep.14.3.159}
}

@book{binmore2007playing,
title     = {Playing for Real: A Text on Game Theory},
author    = {Binmore, Ken},
year      = {2007},
publisher = {Oxford University Press},
address   = {Oxford}
}

@article{oudah24,
author = {Mayada Oudah and Kinga Makovi and Kurt Gray and Balaraju Battu and Talal Rahan},
title = {Perception of experience influences altruism and perception of agency influences trust in human–machine interactions},
journal = {Scientific Reports} ,
volume = {14},
number = {12410},
year = {2024}
}

@ARTICLE{uncanny,
author={Mori, Masahiro and MacDorman, Karl F. and Kageki, Norri},
journal={IEEE Robotics \& Automation Magazine}, 
title={The Uncanny Valley [From the Field]}, 
year={2012},
volume={19},
number={2},
pages={98-100},
doi={10.1109/MRA.2012.2192811}}

@ARTICLE{Sagliano2022,
AUTHOR={Sagliano, Laura  and Ponari, Marta  and Conson, Massimiliano  and Trojano, Luigi },
TITLE={Editorial: The interpersonal effects of emotions: The influence of facial expressions on social interactions},
JOURNAL={Frontiers in Psychology},
VOLUME={13},
YEAR={2022},
DOI={10.3389/fpsyg.2022.1074216},
ISSN={1664-1078},

}

@article{contempt,
author = {Matsumoto, David and Ekman, Paul},
year = {2004},
month = {10},
pages = {529-40},
title = {The Relationship Among Expressions, Labels, and Descriptions of Contempt},
volume = {87},
journal = {Journal of personality and social psychology},
doi = {10.1037/0022-3514.87.4.529}
}

@ARTICLE{iMotionsAccuracy,
AUTHOR={Kulke, Louisa  and Feyerabend, Dennis  and Schacht, Annekathrin },       
TITLE={A Comparison of the Affectiva iMotions Facial Expression Analysis Software With EMG for Identifying Facial Expressions of Emotion},
JOURNAL={Frontiers in Psychology},
VOLUME={11},
YEAR={2020},
DOI={10.3389/fpsyg.2020.00329},
ISSN={1664-1078}
}

@article{Stckli2018FacialEA,
title={Facial expression analysis with AFFDEX and FACET: A validation study},
author={Sabrina St{\"o}ckli and Michael Schulte-Mecklenbeck and Stefan Borer and Andrea C. Samson},
journal={Behavior Research Methods},
year={2018},
volume={50},
pages={1446-1460}
}

@article{CENTORRINO20158,
title = {Honest signaling in trust interactions: smiles rated as genuine induce trust and signal higher earning opportunities},
journal = {Evolution and Human Behavior},
volume = {36},
number = {1},
pages = {8-16},
year = {2015},
issn = {1090-5138},
doi = {10.1016/j.evolhumbehav.2014.08.001},
author = {Samuele Centorrino and Elodie Djemai and Astrid Hopfensitz and Manfred Milinski and Paul Seabright},
}

@book{knapp,
author = {Mark L. Knapp and Judith A. Hall and Terrence G. Horgan},
title = {Nonverbal Communication in Human Interaction (8th ed.)},
publisher = {Boston, MA: Cengage Learning},
year = {2013}
}

@article{role2009,
ISSN = {09637214},
author = {Chris Frith},
journal = {Philosophical Transactions of The Royal Society Biological Sciences},
volume = {364},
pages = {3453–3458},
title = {Role of facial expressions in social interactions},
doi = {10.1098/rstb.2009.0142},
year = {2009}
}

@article{Kleef2022,
author = "van Kleef, Gerben A. and Côté, Stéphane",
title = "The Social Effects of Emotions", 
journal= "Annual Review of Psychology",
year = "2022",
volume = "73",
number = "Volume 73, 2022",
pages = "629-658",
doi = "10.1146/annurev-psych-020821-010855",
publisher = "Annual Reviews",
issn = "1545-2085",
type = "Journal Article",
}

@article{facialcues2010,
author = {M. Stirrat and D.I. Perrett},
title ={Valid Facial Cues to Cooperation and Trust: Male Facial Width and Trustworthiness},

journal = {Psychological Science},
volume = {21},
number = {3},
pages = {349-354},
year = {2010},
doi = {10.1177/0956797610362647}
}

@article{Holland_2020,
author = {Alison C. Holland and Garret O’Connell and Isabel Dziobek},
title = {Facial mimicry, empathy, and emotion recognition: a meta-analysis of correlations},
journal = {Cognition and Emotion},
volume = {35},
number = {1},
pages = {150--168},
year = {2021},
publisher = {Routledge},
doi = {10.1080/02699931.2020.1815655}
}

@article{Kleef_2009,
author = {Gerben A. Van Kleef},
title ={How Emotions Regulate Social Life: The Emotions as Social Information (EASI) Model},
journal = {Current Directions in Psychological Science},
volume = {18},
number = {3},
pages = {184-188},
year = {2009},
doi = {10.1111/j.1467-8721.2009.01633.x}
}

@book{Ekman2003,
author = {Paul Ekman},
title = {Emotions revealed: Recognizing faces and feelings to improve communication and emotional life},
publisher = {Times Books/Henry Holt and Co.},
year = {2003}
}

@article{Hatfield1993,
 ISSN = {09637214},
 author = {Elaine Hatfield and John T. Cacioppo and Richard L. Rapson},
 journal = {Current Directions in Psychological Science},
 number = {3},
 pages = {96--99},
 publisher = {[Association for Psychological Science, Sage Publications, Inc.]},
 title = {Emotional Contagion},
 urldate = {2024-10-08},
 volume = {2},
 year = {1993}
}

@article{Blake2015,
title = {The shadow of the future promotes cooperation in a repeated prisoner’s dilemma for children},
journal = {Scientific Reports},
volume = {5},
number = {14559},
year = {2015},
doi = {10.1038/srep14559},
author = {Peter R. Blake and David G. Rand and Dustin Tingley and Felix Warneken}}

@article{Ekman,
title = {An argument for basic emotions},
journal = {Cognition and Emotion},
volume = {6},
number = {3-4},
pages = {169-200},
year  = {1992},
publisher = {Routledge},
doi = {10.1080/02699939208411068},
author = {Paul Ekman}}

@article{CHEN201265,
title = {Electrophysiological correlates of processing facial attractiveness and its influence on cooperative behavior},
author = {Jie Chen and Jun Zhong and Youxue Zhang and Peng Li and Aiqun Zhang and Qianbao Tan and Hong Li},
journal = {Neuroscience Letters},
volume = {517},
number = {2},
pages = {65-70},
year = {2012},
issn = {0304-3940},
doi = {10.1016/j.neulet.2012.02.082}
}

@article{Stirrat2010,
author = {M. Stirrat and D.I. Perrett},
title ={Valid Facial Cues to Cooperation and Trust: Male Facial Width and Trustworthiness},
journal = {Psychological Science},
volume = {21},
number = {3},
pages = {349-354},
year = {2010},
doi = {10.1177/0956797610362647}
}

@article{Eckel2006,
author = {Rick K. Wilson and Catherine C. Eckel},
title ={Judging a Book by its Cover: Beauty and Expectations in the Trust Game},
journal = {Political Research Quarterly},
volume = {59},
number = {2},
pages = {189-202},
year = {2006},
doi = {10.1177/106591290605900202}
}

@manual{iMotions, 
title={iMotions Software, Version 9.0}, 
url={https://imotions.com/}, 
author={{iMotions A/S}}, 
year={2026}, month={January}}

@ARTICLE{deMelo2015,
author={de Melo, Celso M. and Gratch, Jonathan and Carnevale, Peter J.},
journal={IEEE Transactions on Affective Computing}, 
title={Humans versus Computers: Impact of Emotion Expressions on People's Decision Making}, 
year={2015},
volume={6},
number={2},
pages={127-136},
doi={10.1109/TAFFC.2014.2332471}}

@article{deMelo20,
title = {The interplay of emotion expressions and strategy in promoting cooperation in the iterated prisoner’s dilemma},
journal = {Scientific Reports},
volume = {10},
number ={14959},
year = {2020},
doi = {10.1038/s41598-020-71919-6},
author = {Celso M. de Melo and Kazunori Terada}
}

@article{Reuben2009,
author = {Hopfensitz, Astrid and Reuben, Ernesto},
title = "{The Importance of Emotions for the Effectiveness of Social Punishment}",
journal = {The Economic Journal},
volume = {119},
number = {540},
pages = {1534-1559},
year = {2009},
month = {07},
issn = {0013-0133},
doi = {10.1111/j.1468-0297.2009.02288.x}
}

@article{Mussel2013, 
title={The value of a smile: Facial expression affects ultimatum-game responses}, 
volume={8}, 
DOI={10.1017/S1930297500006045}, 
number={3}, 
journal={Judgment and Decision Making}, 
author={Mussel, Patrick and Göritz, Anja S. and Hewig, Johannes}, 
year={2013}, 
pages={381–385}}
\bibliographystyle{naturemag}

\section*{Acknowledgments}
The authors are deeply grateful to Professor Rebecca Morton (1954--2020) for her early and foundational contribution to the conception of this study. Although she passed away before the final study was implemented, her ideas and guidance were pivotal in shaping its early direction. 

This research is supported by ASPIRE Award for Research Excellence (AARE-2019-48). Wooders gratefully acknowledges financial support from Tamkeen under the NYU Abu Dhabi Research Institute Award CG005. 

\vspace*{-5mm}

\section*{Author contributions statement}
M.O. conceived the study. M.O. and J.W. designed the experiments. M.O. conducted the experiments and collected the data. M.O. and J.W. analyzed the data, interpreted the results, and wrote the manuscript.

\section*{Competing interests statement}
The authors declare no competing interests.

\clearpage
\newpage


\section*{Supplementary material}

# Supplementary Information - Real-time Facial Communication Restores Cooperation After Defection in Social Dilemmas

Mayada Oudah[1,*] and John Wooders[1,2]

[1]Social Science Division, New York University Abu Dhabi, UAE.
[2]Center for Behavioral Institutional Design, New York University Abu Dhabi, UAE.

[*]Correspondence should be addressed to mayada@nyu.edu

## Contents

1. Supplementary Methods

2. Supplementary Figures

3. Supplementary Tables

# 1   Supplementary Methods

**Experimental Design and Payoffs**

Participants played an interactive version of the repeated Prisoner's Dilemma with an indefinite horizon (random termination of 10% after 30 rounds). Figures S1 to S5 show the graphical user interface of the game and all possible outcomes. In every round, each participant chose between Cooperation (Pushing the lever) or Defection (Pulling the lever). The payoffs (in experimental points) were determined by the standard matrix structure: $T > R > P > S$ and $2R > T + S$.

The specific parameters used in the experiment were:

- Reward (R): 60 points (Mutual Cooperation)

- Temptation (T): 80 points (Defecting against a Cooperator)

- Punishment (P): 20 points (Mutual Defection)

- Sucker (S): 0 points (Cooperating against a Defector)

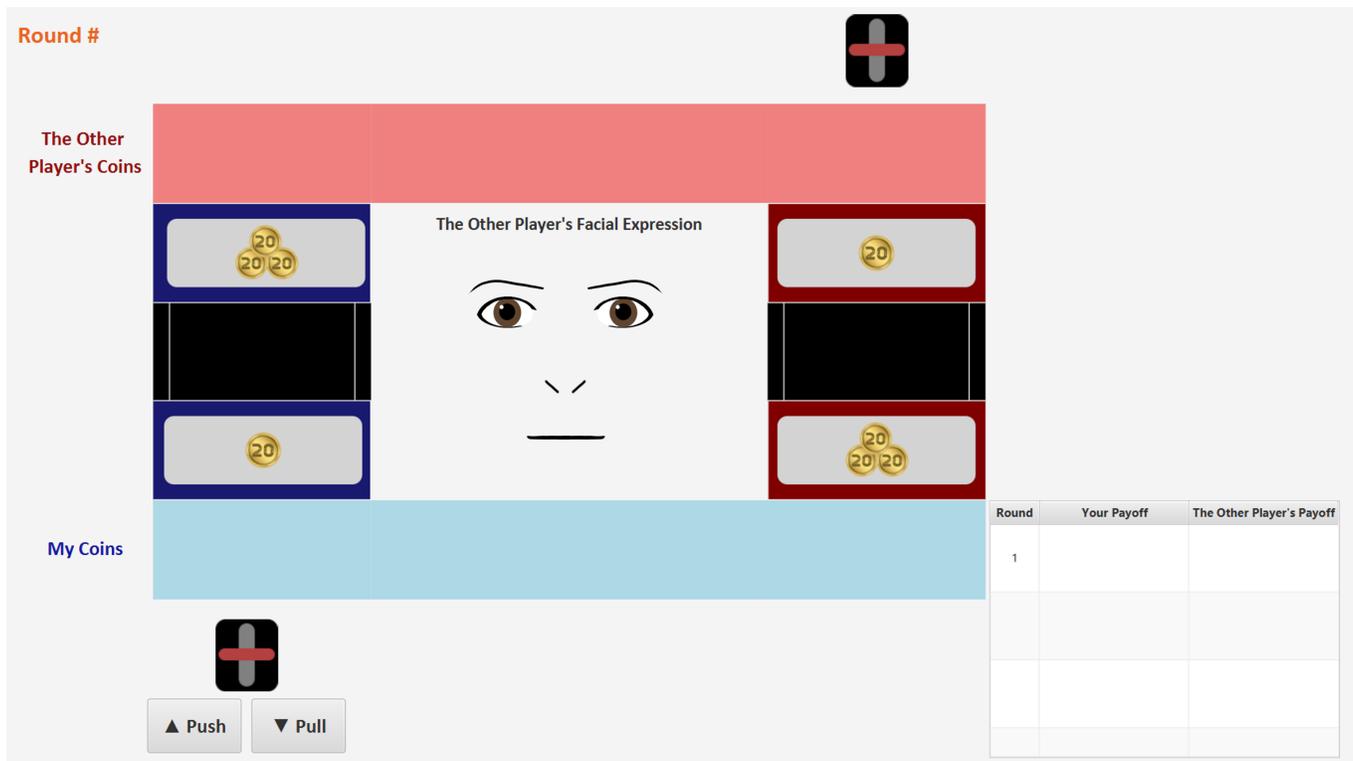

Figure S1: **Game Graphical User Interface in the FC condition.** Participants see themselves as the player on the bottom left side with the other player on the opposite right end. Each player can either push (Cooperate) or pull (Defect) their tray via a lever by clicking the corresponding buttons for these actions. Pushing the tray delivers three coins to the other player and causes one coin to fall into a black hole. Pulling the tray delivers one coin to the player and causes three coins to fall into the black hole. The middle block is blank in the nFC condition.

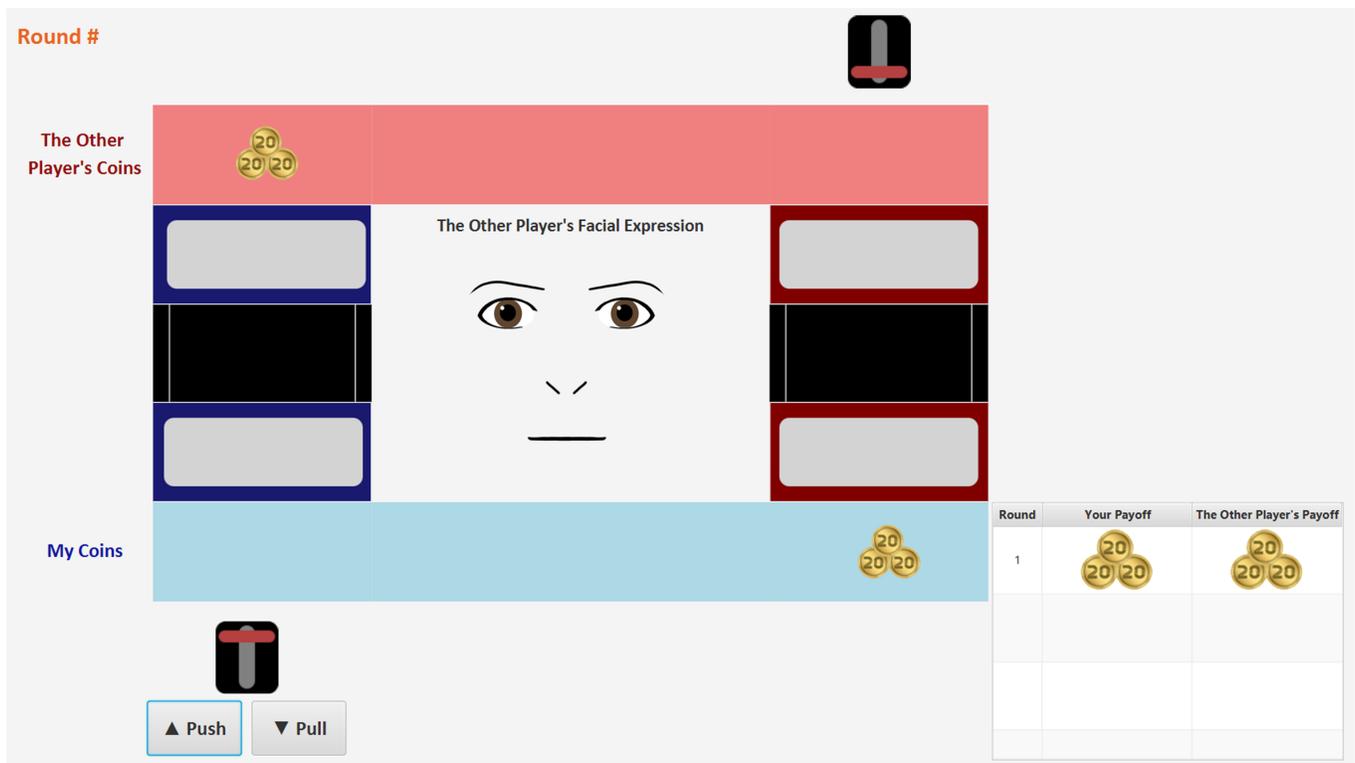

Figure S2: GUI: Mutual cooperation.

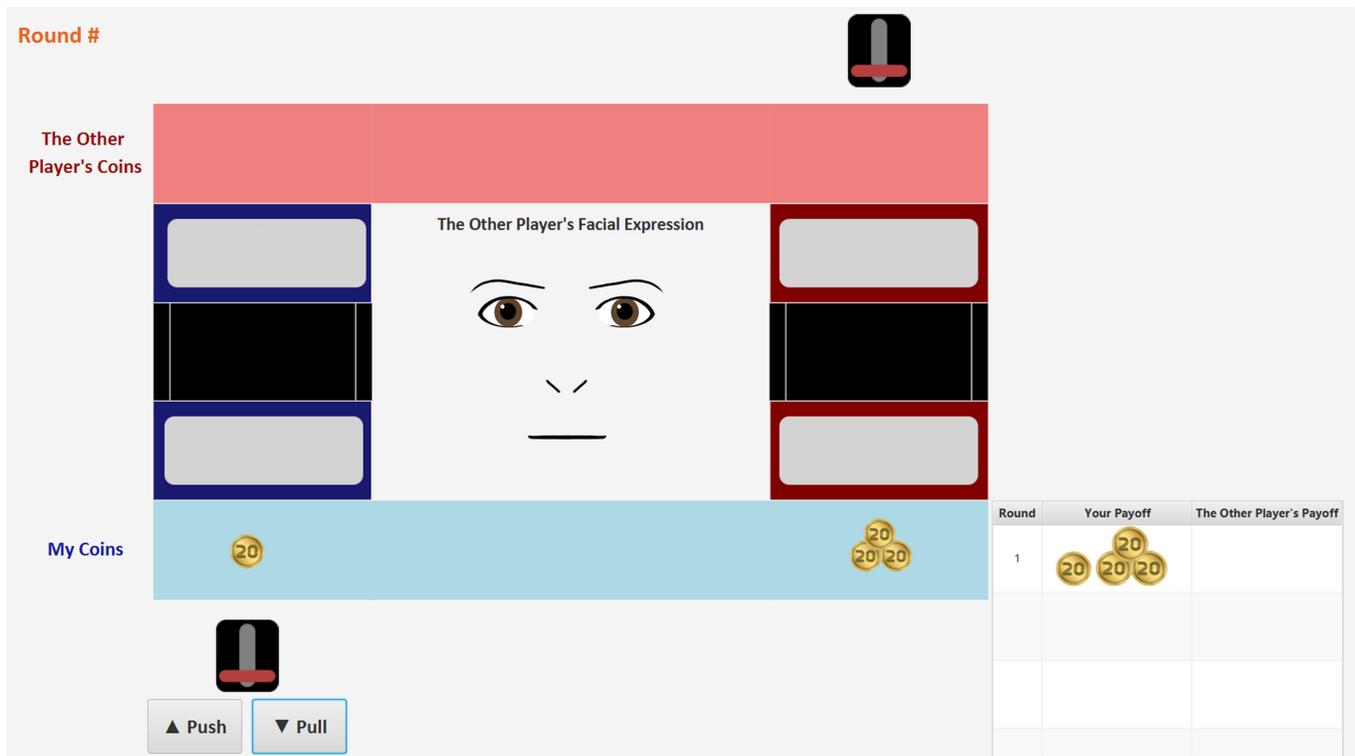

Figure S3: GUI: Defecting against a cooperator.

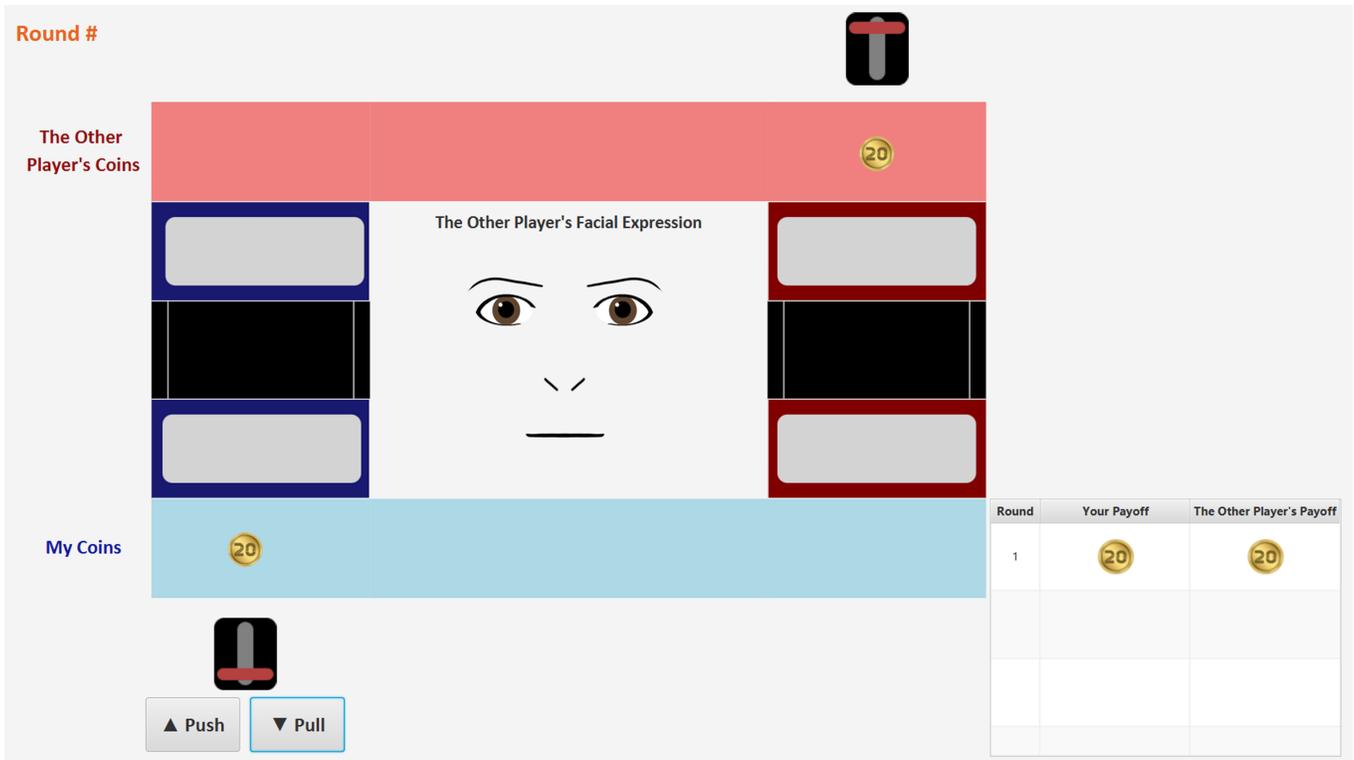

Figure S4: GUI: Mutual defection.

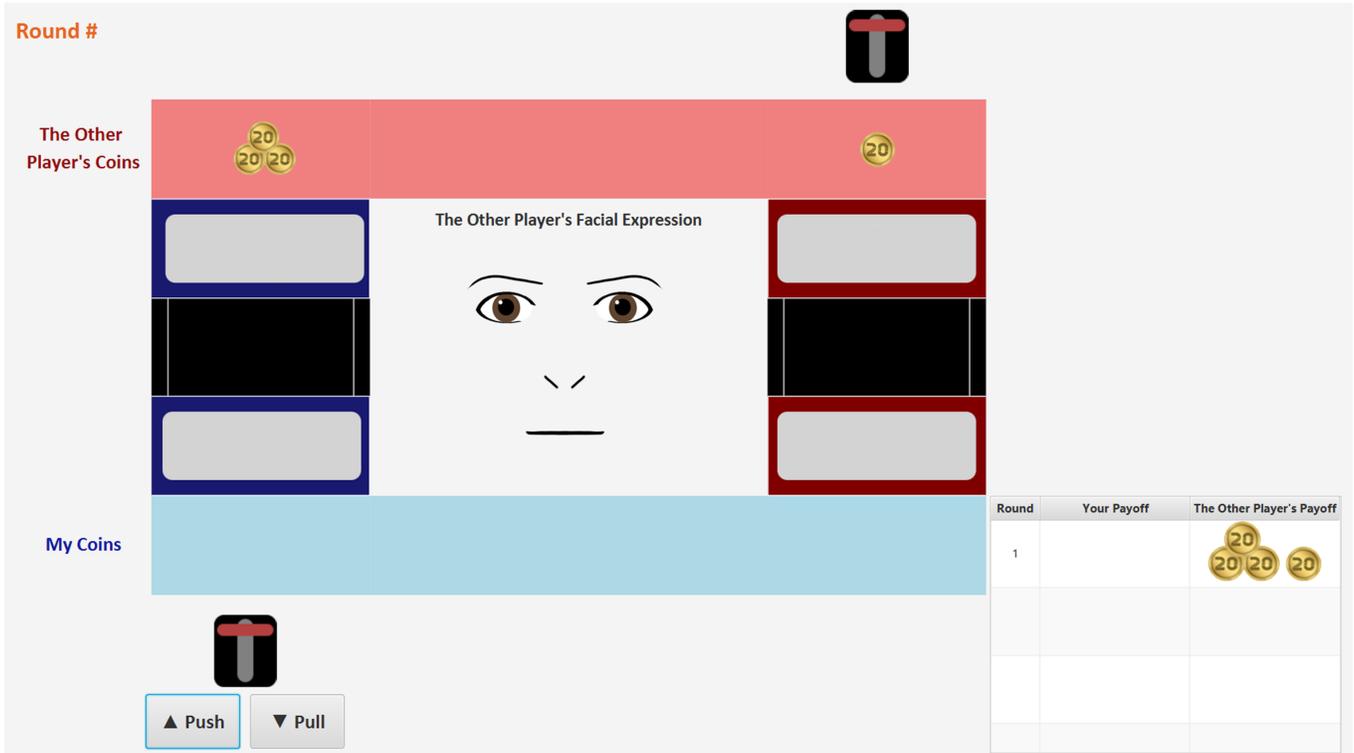

Figure S5: GUI: Cooperating against a defector.

## Biometric Capture and Classification Specifications

Participants' facial expressions were captured and analyzed in real-time using the iMotions Biometric Research Platform (Version 9.0), which integrates the Affectiva AFFDEX emotion recognition algorithm. The AFFDEX engine utilizes a deep learning-based computer vision pipeline to detect facial landmarks and classify expressions based on the Facial Action Coding System (FACS) established by Ekman and Friesen [1].

The analysis proceeds in three stages. First, the algorithm isolates the participant's face from the video feed using a Histogram of Oriented Gradients (HOG) and Support Vector Machine (SVM) classifier. Second, it maps 34 distinct facial landmarks (including eye corners, nose tip, and mouth contours) to model the geometric configuration of the face . Finally, the relationship between these landmarks is analyzed to detect specific Action Units (AUs) – the fundamental building blocks of expression (e.g., AU12, "Lip Corner Puller"). These AUs are aggregated via a pre-trained Deep Learning model to generate probability scores (0–100) for Neutral and seven core emotional states: Joy, Anger, Surprise, Fear, Contempt, Disgust, and Sadness.

The Affectiva AFFDEX architecture was trained on a large-scale naturalistic dataset comprising millions of facial frames, ensuring high robustness across diverse demographics, lighting conditions, and head poses (McDuff et al. [2]). Its classification accuracy has been further validated in independent studies against human expert coders. Stöckli et al. [3] reported that the software achieves concordance rates comparable to human coders for distinct, high-intensity expressions, particularly for Joy (Accuracy $> 90\%$). To ensure high data quality in our specific experimental setup, we maintained controlled lighting conditions and fixed participant positioning to minimize extreme head-pose variations that could degrade classifier performance.

Figure S6 shows iMotions module capturing a neutral expression. Neutral expression usually entails no valence and low (or no) engagement. Figure S7 shows the module capturing a joyful expression. Positive expressions lead to positive valence and high engagement, whereas negative expressions lead to negative valence and high engagement if the participant is focusing on the screen.

## Displaying and Analyzing Emotions

In the Facial Communication treatment, a participant's facial expression is communicated, via an avatar, dynamically in real-time to the participant they are matched with. At any moment, the avatar displayed corresponds to the emotion with the highest intensity value, where each emotion is quantified on a 0–100 scale. A threshold of 60% is imposed such that, if no emotion exceeds this value at a given moment, a Neutral avatar is shown. If the software fails to detect a face (e.g., the participant looked away from the screen), no avatar is communicated.

In our post-hoc regression analyses, we utilized the raw, continuous probability scores (0-100) rather than the thresholded binary states. This allowed us to detect subtle emotional shifts and "pre-emotions" (e.g., a rise in Joy from 15 to 50) that indicated internal affective engagement even if they were not intense enough to trigger the avatar display.

To maximize the accuracy and efficiency of facial expression analysis, we annotated key events within each interaction with timestamps. These events included the beginning of a round, the moment an action was selected, the observation of the joint action, the end of a round, and the termination of the interaction. For each round, the timeline was divided into two intervals: one preceding the participant's action selection and one following it. We excluded from the analysis those times during which the engagement level fell below 30%. Engagement reflects the participant's visual attention toward the camera, which was positioned

directly above the monitor used for viewing information and making decisions. Low engagement indicates that the participant was looking away from the screen or covering their face. For each round, each emotion is summarized by two means: one for the measured pre-action and one for the measured post-action intensity values.

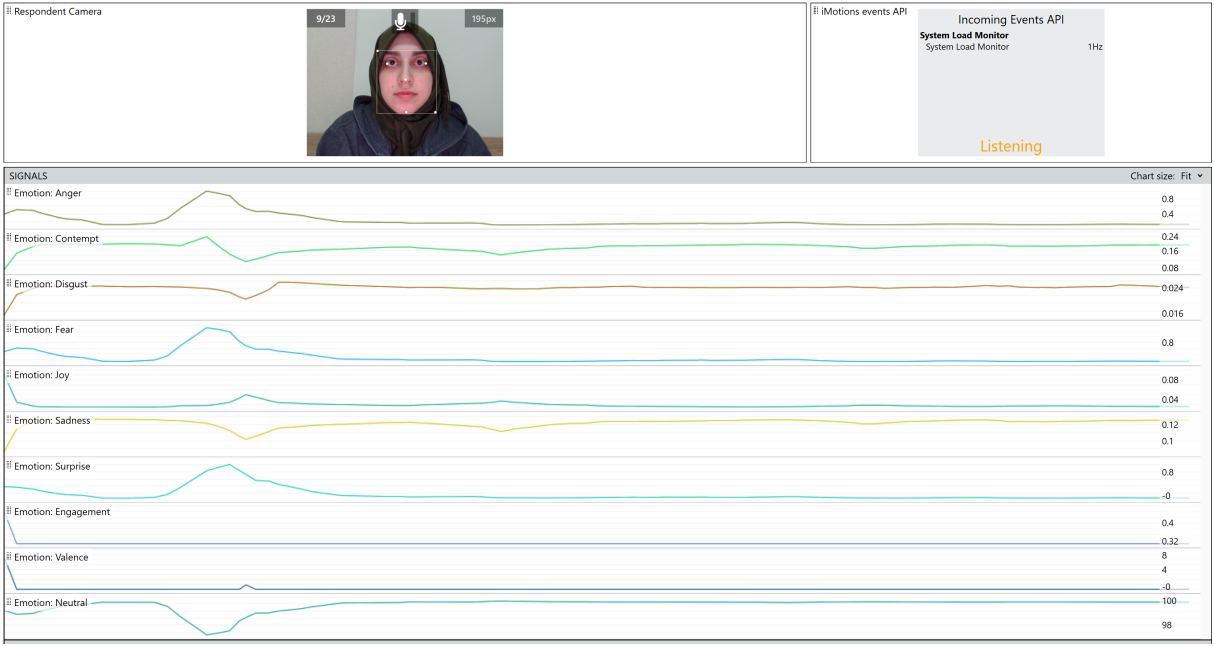

Figure S6: iMotions Facial Expression Analysis Module – Neutral Expression.

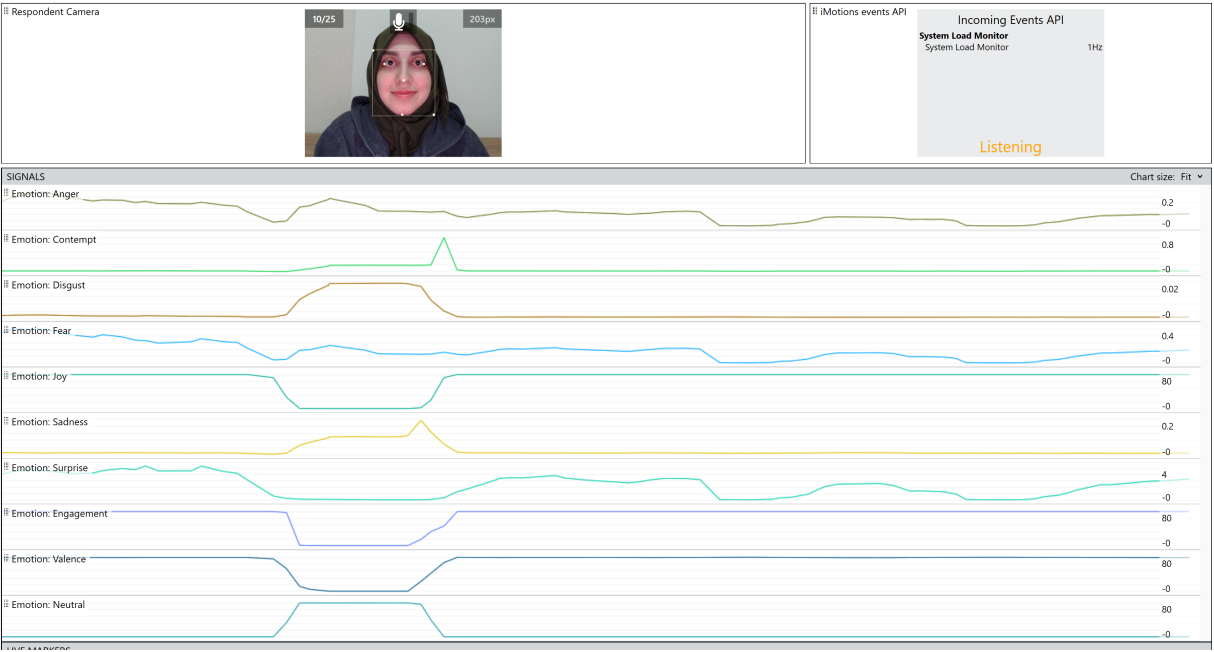

Figure S7: iMotions Facial Expression Analysis Module – Joy Expression.

## Avatar Validation and Recognition Accuracy

To ensure our gender-neutral avatars accurately conveyed the intended emotional signals, we conducted an independent validation study prior to the main experiment. Participants ($N = 113$) were presented with our designed avatars and asked to classify them into one of the eight target emotion categories. Figure S8 presents the confusion matrix of these classifications. The diagonal elements represent the "Hit Rate" (correct classification).

Crucially for our main findings, the recognition rate for Joy—the primary signal associated with the restoration of cooperation—was extremely high ( $99\%$). This confirms that the "forgiveness" mechanism observed in the main text was driven by an unambiguous positive signal.

As is common in facial expression research [4, 5, 6], recognition rates were lower for high-nuance negative emotions, specifically Contempt ( $57\%$). However, the off-diagonal errors in the matrix reveal that misclassifications were largely restricted to other negative-valence categories (e.g., confusing Contempt with Disgust or Anger and assigning it as "Hard to Classify"). Consequently, while the specific label of the negative emotion may have varied, the directional signal (social disapproval/negativity) remained preserved. This suggests that even if the specific "Contempt" label was not always retrieved, the avatar successfully communicated the partner's dissatisfaction.

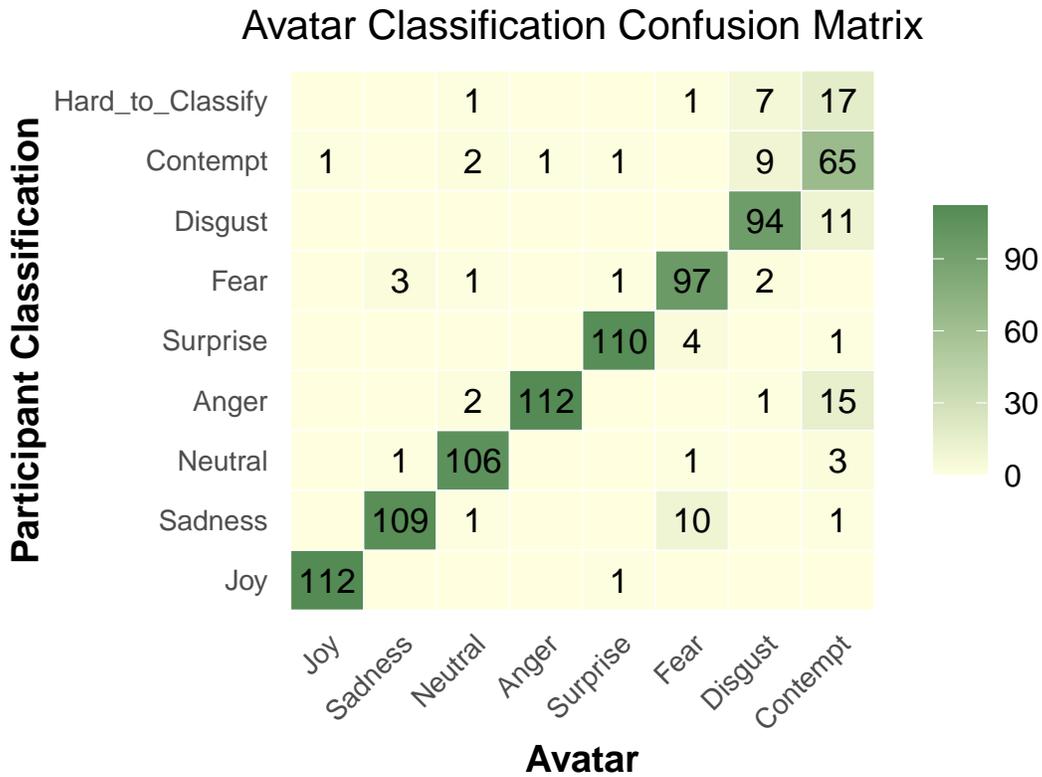

Figure S8: Avatar Classification Confusion Matrix

# 2    Supplementary Figures

Figures S9 to S13 show the survey participants took to evaluate the designed avatars. Figures S14 to S17 show the survey taken by participants post completing the interaction.

## Avatar Evaluation Survey

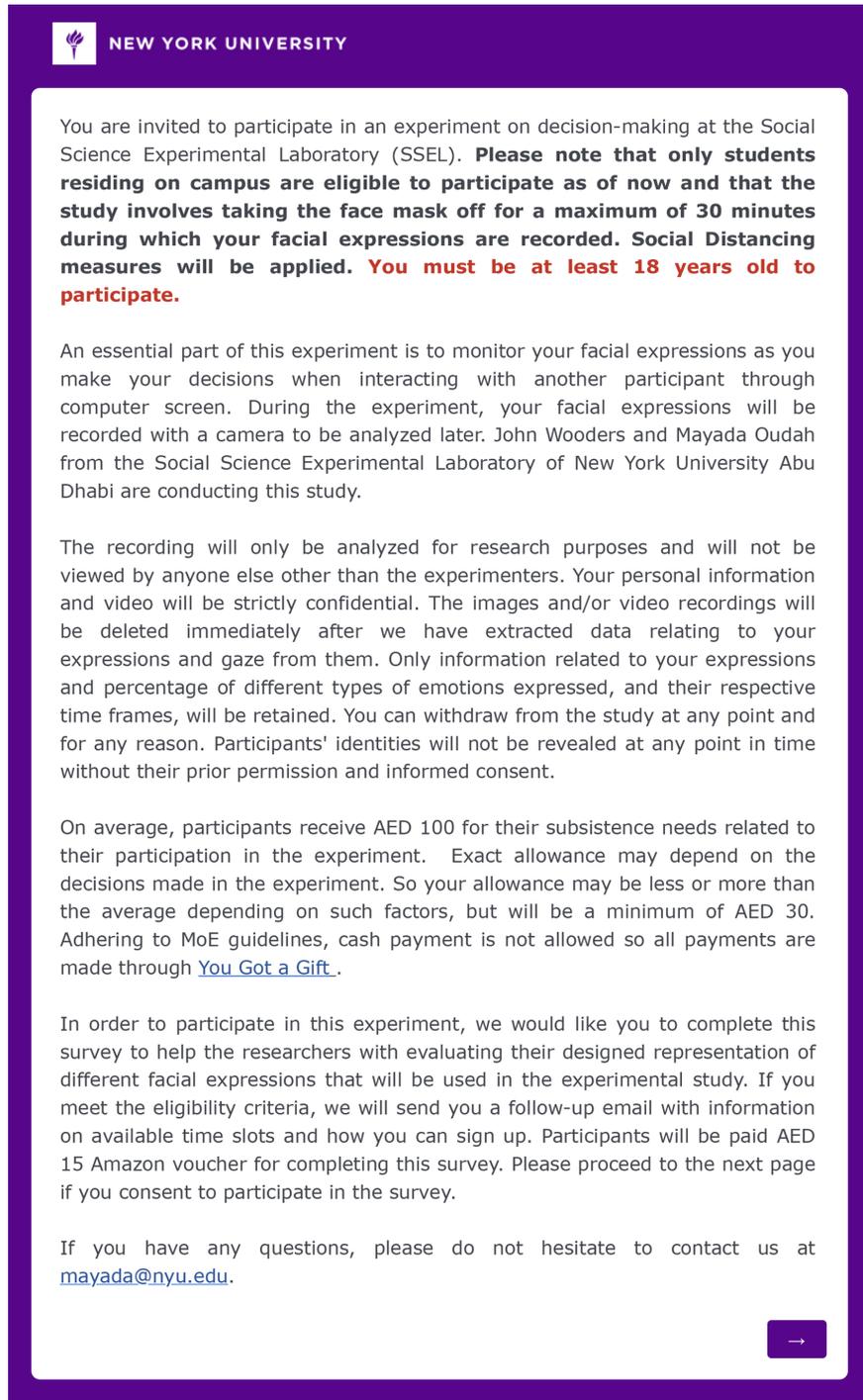

**NEW YORK UNIVERSITY**

You are invited to participate in an experiment on decision-making at the Social Science Experimental Laboratory (SSEL). **Please note that only students residing on campus are eligible to participate as of now and that the study involves taking the face mask off for a maximum of 30 minutes during which your facial expressions are recorded. Social Distancing measures will be applied. You must be at least 18 years old to participate.**

An essential part of this experiment is to monitor your facial expressions as you make your decisions when interacting with another participant through computer screen. During the experiment, your facial expressions will be recorded with a camera to be analyzed later. John Wooders and Mayada Oudah from the Social Science Experimental Laboratory of New York University Abu Dhabi are conducting this study.

The recording will only be analyzed for research purposes and will not be viewed by anyone else other than the experimenters. Your personal information and video will be strictly confidential. The images and/or video recordings will be deleted immediately after we have extracted data relating to your expressions and gaze from them. Only information related to your expressions and percentage of different types of emotions expressed, and their respective time frames, will be retained. You can withdraw from the study at any point and for any reason. Participants' identities will not be revealed at any point in time without their prior permission and informed consent.

On average, participants receive AED 100 for their subsistence needs related to their participation in the experiment. Exact allowance may depend on the decisions made in the experiment. So your allowance may be less or more than the average depending on such factors, but will be a minimum of AED 30. Adhering to MoE guidelines, cash payment is not allowed so all payments are made through You Got a Gift .

In order to participate in this experiment, we would like you to complete this survey to help the researchers with evaluating their designed representation of different facial expressions that will be used in the experimental study. If you meet the eligibility criteria, we will send you a follow-up email with information on available time slots and how you can sign up. Participants will be paid AED 15 Amazon voucher for completing this survey. Please proceed to the next page if you consent to participate in the survey.

If you have any questions, please do not hesitate to contact us at mayada@nyu.edu.

→

Figure S9: Introduction to the survey.

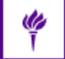 **NEW YORK UNIVERSITY**

**What is your year of birth?**

Year: [ ▾ ]

---

**What is your Gender?**

Male

Female

---

**Which region are you from? Use [this map](#) for help**

East Asia and Pacific

Sub-Saharan Africa

South Asia

Europe and Central Asia

Middle East and North Africa

Latin America and Caribbean

North America

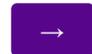

Figure S10: Demographics questions.

9

How would you classify the following expressions? Drag and drop each expression where it fits. Note that you can drop more than one expression under the same category.

**Items**

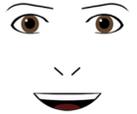
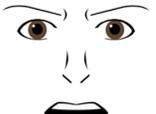
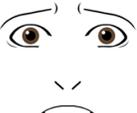
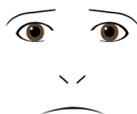
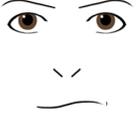
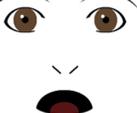
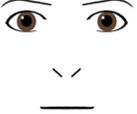
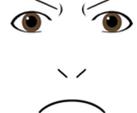
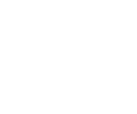

| Joy | Sadness |
|-----|---------|
|     |         |

| Neutral | Anger |
|---------|-------|
|         |       |

| Surprise | Fear |
|----------|------|
|          |      |

| Disgust | Contempt |
|---------|----------|
|         |          |

| Hard to Classify |
|------------------|
|                  |

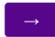

Figure S11: Avatar classification.

**NEW YORK UNIVERSITY**

What gender would you assign the following expressions? Drag and drop each expression where it fits. Note that you can drop more than one expression under the same category.

Items

| Male | Female |
|------|--------|
|      |        |

Can be either

Avatar facial expressions (eyes and mouths)

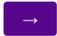

Figure S13: Avatar accuracy question.

# Post Experiment Survey

Thank you for taking part in this study!  Please answer all of the questions below

**What is your player ID? (the number on the sticker you were handed)**

**How old are you?**

**What is your Gender?**

○ Male
○ Female
○ Other

**Which region are you from? Use this map for help**

○ East Asia and Pacific
○ Sub-Saharan Africa
○ South Asia
○ Europe and Central Asia
○ Middle East and North Africa
○ Latin America and Caribbean
○ North America

**Where have you lived the longest? Use this map for help**

○ East Asia and Pacific
○ Sub-Saharan Africa
○ South Asia
○ Europe and Central Asia
○ Middle East and North Africa
○ Latin America and Caribbean
○ North America

Figure S14: Demographics questions in the post-experiment survey.

**On the scale 0–7, to what extent do the following terms describe your match behavior in the game?**

**How likable was the other player?**

Not at all                                                                                          Always
0          1          2          3          4          5          6          7

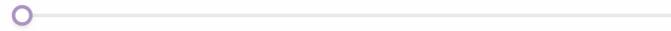

**How intelligent did the other player act?**

Not at all                                                                                          Always
0          1          2          3          4          5          6          7

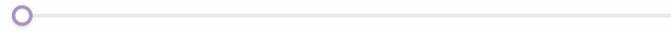

**How cooperative was the other player?**

Not at all                                                                                          Always
0          1          2          3          4          5          6          7

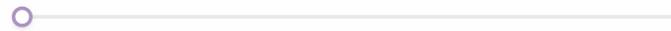

**How trustworthy was the other player?**

Not at all                                                                                          Always
0          1          2          3          4          5          6          7

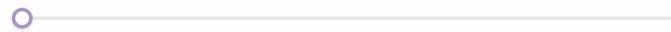

Figure S15: Rating the other player's likability, intelligence, cooperativeness, and trustworthiness.

**How selfish was the other player?**

Not at all                                                                          Always
0           1           2           3           4           5           6           7

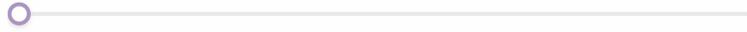

**How forgiving was the other player?**

Not at all                                                                          Always
0           1           2           3           4           5           6           7

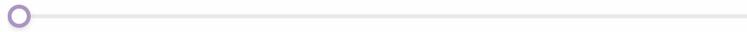

**Was the other player a bully?**

Not at all                                                                          Always
0           1           2           3           4           5           6           7

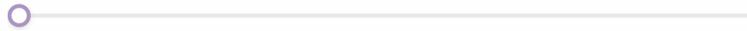

Figure S16: Rating the other player's selfishness, forgivingness and being a bully.

**If given the opportunity, would you want to interact with the other player again?**

○ Yes
○ No

**Why did you make the decisions you made? Describe your strategy for choosing whether to push or pull.**

Figure S17: The "Interact again" question and understanding the participant's strategy for making a decision.

# 3 Supplementary Tables

|  | Cooperation |
|---|---|
| Intercept | 0.69 (0.16)*** |
| Treatment (FC) | 0.47 (0.07)*** |
| AIC | 5670.99 |
| BIC | 5690.63 |
| Log Likelihood | -2832.49 |
| Num. obs. | 5160 |
| Num. groups: PlayerID | 86 |
| Var: PlayerID (Intercept) | 2.04 |

***$p < 0.001$; **$p < 0.01$; *$p < 0.05$

Table S1: Regression results for Cooperation by Treatment

|  | Cooperation | Cooperation | Cooperation |
|---|---|---|---|
| Intercept | -2.00 (0.12)*** | 0.59 (0.16)*** | -2.32 (0.91)* |
| Treatment (FC) | 0.19 (0.09)* | 0.27 (0.07)*** | 0.14 (0.10) |
| Pre-Joy (Own) |  | 0.64 (0.09)*** | 0.23 (0.65) |
| Pre-Joy (Other) |  | 0.21 (0.09)* | 0.31 (0.68) |
| Previous Cooperation (Own) | 1.69 (0.09)*** |  | 1.69 (0.09)*** |
| Previous Cooperation (Other) | 2.76 (0.09)*** |  | 2.74 (0.09)*** |
| Pre-Contempt (Own) |  |  | -0.07 (0.68) |
| Pre-Disgust (Own) |  |  | -0.66 (0.92) |
| Pre-Fear (Own) |  |  | 0.37 (1.08) |
| Pre-Neutral (Own) |  |  | -0.04 (0.65) |
| Pre-Sadness (Own) |  |  | 0.40 (0.79) |
| Pre-Surprise (Own) |  |  | -1.08 (0.90) |
| Pre-Contempt (Other) |  |  | 0.48 (0.72) |
| Pre-Disgust (Other) |  |  | -0.82 (0.99) |
| Pre-Fear (Other) |  |  | 0.05 (1.11) |
| Pre-Neutral (Other) |  |  | 0.36 (0.68) |
| Pre-Sadness (Other) |  |  | 0.84 (0.83) |
| Pre-Surprise (Other) |  |  | -0.91 (0.88) |
| AIC | 3580.65 | 5617.29 | 3589.35 |
| BIC | 3613.22 | 5650.03 | 3713.13 |
| Log Likelihood | -1785.32 | -2803.64 | -1775.67 |
| Num. obs. | 4988 | 5160 | 4988 |
| Num. groups: PlayerID | 86 | 86 | 86 |
| Var: PlayerID (Intercept) | 0.38 | 2.04 | 0.38 |

***$p < 0.001$; **$p < 0.01$; *$p < 0.05$

Table S2: Regression results for Cooperation by Treatment, Previous Actions and Previous Emotions

# References

[1] Ekman, P. & Friesen, W. Constants across cultures in the face and emotion. Journal of Personality and Social Psychology **17**, 124–129 (1971).

[2] McDuff, D. et al. Affdex sdk: A cross-platform real-time multi-face expression recognition toolkit. In Proceedings of the 2016 CHI Conference Extended Abstracts on Human Factors in Computing Systems, 3723–3726 (Association for Computing Machinery, New York, NY, USA, 2016).

[3] Stöckli, S., Schulte-Mecklenbeck, M., Borer, S. & Samson, A. C. Facial expression analysis with affdex and facet: A validation study. Behavior Research Methods **50**, 1446–1460 (2018). URL `https://api.semanticscholar.org/CorpusID:27426298`.

[4] Matsumoto, D. & Ekman, P. The relationship among expressions, labels, and descriptions of contempt. Journal of personality and social psychology **87**, 529–40 (2004).

[5] Russell, J. A. The contempt expression and the relativity thesis. Motivation and Emotion **15**, 149–168 (1991).

[6] Ekman, P., O'Sullivan, M. & Matsumoto, D. Confusions about context in the judgment of facial expression: A reply to "the contempt expression and the relativity thesis. Motivation and Emotion **15**, 169–176 (1991).


\end{document}